\documentclass[preprint]{revtex4-1}
\usepackage{graphicx}
\usepackage{amsmath}
\usepackage[colorlinks]{hyperref}
\usepackage{lineno}

\newcommand{\tb}{t_\beta}

\newcommand{\cba}{c_{\beta\alpha}}
\newcommand{\sba}{s_{\beta\alpha}}
\newcommand{\maa}{m_{A^0}}
\newcommand{\mhl}{m_{h^0}}
\newcommand{\mhh}{m_{H^0}}
\newcommand{\mhp}{m_{H^\pm}}
\newcommand{\products}{H^+H^-Z^0}
\newcommand{\theprocess}{e^+e^- \rightarrow \products}
\newcommand{\htb}{H^+\rightarrow t \bar{b}}
\newcommand{\hwh}{H^+\rightarrow W^+h^0}
\newcommand{\hwH}{H^+\rightarrow W^+H^0}

\newcommand{\fb}{\text{ fb}}
\newcommand{\gev}{\text{ GeV}}
\newcommand{\tev}{\text{ TeV}}
\newcommand{\factor}{0.40}
\newcommand{\factortwo}{0.32}

\begin{document}

\title{Charged Higgs pair production in association with the $Z^0$ boson at electron-positron colliders}

\author{Nasuf SONMEZ}
\email{nasuf.sonmez@ege.edu.tr}
\affiliation{Department of Physics, Faculty of Science, Ege University, 35040, Izmir}


\date{\today}

\begin{abstract}
    In this paper, the production of the charged Higgs pair associated with the $Z^0$ boson is analyzed in the minimal extension of the standard model the so-called two-Higgs-doublet model (2HDM). The process $e^+e^- \rightarrow H^+H^-Z^0$ is calculated at the tree level including all the possible diagrams in 2HDM. The numerical analysis is performed in consideration of the current experimental constraints and various scenarios for the free parameters of the model. The results are presented as a function of center-of-mass energy, the charged Higgs mass ($m_{H^\pm}$), and the ratio of the vacuum expectation values ($t_\beta$). The unpolarized cross section, taking into account the results in the flavor physics, gets up to $0.278\text{ fb}$ for $m_{H^\pm}=175\text{ GeV}$, and it declines with decreasing $m_{H^\pm}$ in Type-I. However, it gets down to $0.073\text{ fb}$ for $m_{H^\pm}=500\text{ GeV}$ in Type-II. Further, the calculation is also carried out in the non-alignment scenario and low-$m_{h^0}$ scenario. The effect of the polarized incoming $e^+$ and $e^-$ beams shows that the cross section is enhanced by a factor up to 2.5 at P(+0.60,-0.80) polarization configuration. Decay channels of the charged Higgs, possible final states of the process, and some differential distributions belong to the charged Higgs and $Z^0$ boson are examined for each scenario. The analysis shows that some channels have higher branching ratio such as $H^+\rightarrow t \bar{b}$, $H^+\rightarrow W^+h^0$, and $H^+\rightarrow W^+H^0$. These decay channels are essential for the charged Higgs searches in the lepton colliders regarding the scenarios interested. The detection of the charged Higgs is a powerful sign for the extended scalar sectors, and the results show the potential of a future lepton collider.
\end{abstract}


\maketitle


 \section{Introduction}

    Over the last couple of decades, many extensions to cure the quadratic divergence at the scalar sector of the Standard Model (SM) has been proposed, and the implications of the new physics have been studied intensively. One possible extension of the SM is to add a second Higgs doublet to the scalar sector. These two Higgs doublets defined to have the same quantum number so that they together could give mass to leptons, quarks, and electroweak bosons. Addition of an extra doublet gives new couplings and interactions, in a result rich phenomenology to the 2HDM. In a general 2HDM, there are two charged Higgs bosons $(H^\pm)$ and three neutral Higgs bosons $(h^0, A^0, H^0)$ \cite{Branco:2011iw, Gunion:1989we} playing with the free parameters of the model $h^0$ could be set to resemble the discovered Higgs boson.

    Nowadays, there is an ongoing effort for another project named Linear Collider Collaboration (LCC), it is an organization that brings the two most likely candidates for the next collider program, the Compact Linear Collider Study (CLIC) and the International Linear Collider (ILC), together under one roof. When this project is constructed, there $e^+e^-$, $e^-e^-$ and $\gamma e$ collisions will be studied. One of the primary task at the future lepton colliders is to complement the LHC results, and also searching for clues in beyond the SM such as supersymmetry, an extension of the scalar sector or exotic models. Both of the collider projects are designed to study the properties of the new particles and the interactions they might undergo according to the vast amount of theories. As it is known, the lepton colliders compared to the LHC have a cleaner background and it is possible to extract the new physics signals from the background more easily.

    There has been a long time effort to observe a hint associated with a charged Higgs boson in the past and current experiments. However, it was not discovered at the LEP, Tevatron, and yet the search is still going on at the LHC \cite{Laurila:2017phk}. The main discovery channel of the charged Higgs boson in 2HDM is through $e^+e^-\rightarrow H^+H^-$ or $e^+e^-\rightarrow H^\pm ff'$ channels. Such signals in 2HDM are studied in Ref. \cite{Kanemura:2014dea, Arhrib:1998gr}. On the other hand, the pair production of the charged Higgs boson associated with gauge boson is also possible at the lepton colliders where this process could be studied with more precision. The process $\theprocess$ is complementary to the discovery of the charged Higgs pair production. The process would also give a way to measure the couplings $c_{H^+H^-h^0}$ and $c_{H^+H^-H^0}$. These couplings along with the other trilinear Higgs couplings will help us to reconstruct the Higgs potential\footnote{It should be noted that the complete reconstruction of the Higgs potential requires the measurement of the quartic Higgs couplings as well.}. The process is investigated before in Ref. \cite{Arhrib:2008jp} for a case study where the triple Higgs couplings are studied, and the diagrams which are sensitive to the triple Higgs couplings are included in the calculation. The same process is investigated in left-right twin Higgs model where doubly charged Higgs pair production with $Z^0$ boson is analyzed \cite{Liu:2010jp}, and a similar process in the Higgs triplet model is examined in Ref. \cite{Shen:2014rpa}. In this work, the production of the $H^+H^-Z^0$ including all the possible diagrams at the born level is calculated. Numerical analysis of the total cross section is performed as a function of the center-of-mass (COM) energy, the charged Higgs mass, and the free parameters of the 2HDM for the benchmark scenarios with various motivations. In addition to these, the results with different beam polarizations are presented. A discussion is carried on the couplings essential in the production of $\products$ and decay channels of the charged Higgs boson for each scenario. A preliminary calculation of the differential cross section as a function of the kinematical properties of the Z and the $H^\pm$ boson is obtained. Each scenario with subsequent decay channels of $H^\pm$ is explored for later Monte Carlo studies. 
 
    The content of this paper is organized as follows. In Sec. \ref{sec:2}, the scalar sector, free parameters of the model and the masses of the Higgs bosons in 2HDM are reviewed. The machinery, the workflow of the analysis, and the kinematics of the scattering process are explained in Sec. \ref{sec:3}. The constraints coming from the experimental results are underlined in Sec \ref{sec:4}. Numerical analysis is performed on three scenarios, which are named as non-alignment scenario, low-$m_H$ mass scenario, and the favored region in light of the flavor physics results. The numerical results for each of the scenario, including the beam polarization are presented thoroughly in Sec \ref{sec:5}. Decay channels of the charged Higgs and identification of the process are discussed along with the differential distributions in Sec \ref{sec:6}. The conclusion is drawn in Sec \ref{sec:7}.
    
\section{Scalar sector and theoretical framework for 2HDM}
\label{sec:2}

    In this section, the phenomenology of the 2HDM, the scalar sector, and the free parameters in the model are presented. The model itself and the detailed introduction of the framework are studied before by many authors, and it is given in Ref. \cite{Haber:2006ue, Davidson:2005cw, Gunion:1989we, Carena:2002es}. Therefore, only a short review of the 2HDM is presented which is relevant to the analysis. 2HDM is constructed by adding a second $SU(2)_L$ Higgs doublet with the same a hypercharge ($Y=1$) to the scalar sector. If we denote the Higgs doublets as
        \begin{equation}
        \Phi_i = 
        \begin{pmatrix}
        \phi_{i}^+  \\
        \frac{1}{\sqrt{2}}[ v_i+\rho_i+i\eta_i]\\ 
        \end{pmatrix},    
           \label{eq:eq1}
        \end{equation}
	with the scalar potential of the 2HDM given in Eq. \ref{eq:eq2}, there will be in total 14 free parameters.
        \begin{widetext}
        \onecolumngrid
        \begin{eqnarray}
        V(\Phi_1,\Phi_2)=&&m_{11}^2 | \Phi_1|^2+m_{22}^2|\Phi_2|^2 - \left[ m_{12}^2   \Phi_1^{\dagger} \Phi_2 +h.c. \right] \nonumber \\ 
        &+& \frac{\lambda_1}{2}| ( \Phi_1^{\dagger} \Phi_1 )^2  + \frac{\lambda_2}{2} ( \Phi_2^{\dagger} \Phi_2 )^2 +\lambda_3 | \Phi_1 |^2 | \Phi_2 |^2 +\lambda_4 | \Phi_1^{\dagger} \Phi_2 |^2 \label{eq:eq2} \\
        &+& \left [  \frac{\lambda_5}{2} (\Phi_1^{\dagger} \Phi_2)^2 + \left ( \lambda_6 (\Phi_1^{\dagger} \Phi_1 ) + \lambda_7 (\Phi_2^{\dagger} \Phi_2 )  \right) \Phi_1^{\dagger} \Phi_2 + h.c.  \right] .\nonumber 
        \end{eqnarray}
    \end{widetext}
	
	In general, the parameters $m_{11}$, $m_{22}$ and $\lambda_{1,2,3,4}$ are real while $m_{12}$ and $\lambda_{5,6,7}$ are complex. Both of the doublets have the same charge assignment, and they could couple to leptons and quarks as in the SM. However, to suppress the CP violation and the flavor changing neutral currents (FCNC) at the tree level, the construction of the model needs to be constrained in some way. Traditionally, a discrete symmetry ($\mathcal{Z}_2$) was introduced which puts restrictions on the most general form of the Higgs scalar potential and the Higgs-fermion interactions \cite{Hall:1981bc, Deshpande:1977rw, Lavoura:1994yu}. Discrete $\mathcal{Z}_2$ symmetry is simply defined as the invariance of the Lagrangian under the interchange of $\Phi_1 \rightarrow \Phi_1$ and $\Phi_2 \rightarrow -\Phi_2$ (in a generic basis). If the discrete $\mathcal{Z}_2$ symmetry is extended to the Yukawa sector, Higgs-fermion (Yukawa) interactions could be written in a couple of four different and independent ways. Since we do not care the CP violation, we set $\lambda_6 =\lambda_7 =0$. Then the complex parameters $m_{12}$ and $\lambda_5$ are taken as real. As a result, the free parameter number reduces to 8 under these assumptions. If this symmetry is allowed to violate softly, then FCNCs are naturally suppressed at the tree level. The $m^2_{12}$ term in Eq. \ref{eq:eq2} ensures the breaking of the discrete symmetry softly.

    The mass spectrum of the Higgs bosons is computed first by imposing the constraints obtained from the potential minimum condition ($\partial V/\partial \Phi_i=0, \text{ for i=1,2}$) and eliminating $m_{11}$ and $m_{22}$. Second, the Higgs doublets are defined as given in detail in Ref. \cite{Haber:1978jt, Branco:2011iw, Haber:1978jt, Haber:2006ue, Asakawa:2008se}. After decomposing the scalar potential into a quadratic term plus cubic and quartic interactions, the mass terms are extracted. Finally, diagonalizing the quadratic terms, we easily obtained the physical Higgs states and their masses. The masses of the charged and the CP-odd Higgs states are defined as follows:
        \begin{eqnarray}
            m^2_{A^0}   &=&\frac{m_{12}^2}{\sin\beta\cos\beta} -2\lambda_5 v^2 \\
           m^2_{H^\pm} &=&m^2_{A^0} + (\lambda_5-\lambda_4)v^2
        \end{eqnarray}
        where $\beta$ is the ratio of the vev of the Higgs doublets ($\tan\beta=v_1/v_2$). The mass of the CP-even states becomes
        \begin{equation}
        \begin{pmatrix}
        h^0 \\
        H^0        
        \end{pmatrix} 
        = \mathcal{R}
        \begin{pmatrix}
        m_{12}^2\tan\beta +\lambda_1 v_1 ^2        &    -m_{12}^2+\lambda_{345} v_1 v_2    \\
        -m_{12}^2+\lambda_{345} v_1 v_2 & m_{12}^2\cot\beta +\lambda_2 v_2^2
        \end{pmatrix} \mathcal{R}^T,
        \end{equation} 
        where $\mathcal{R}$ is a unitary rotation matrix which diagonalizes the CP-neutral Higgs mass matrix \cite{Davidson:2005cw, Branco:2011iw} as a function of the angle $(\beta-\alpha)$, $\lambda_{345}=\lambda_{3}+\lambda_{4}+\lambda_{5}$, and $v_1=v\sin\beta,\;v_2=v\cos\beta$ so $v = \sqrt{v_1^2+v_2^2}=246\gev$. Then, the physical neutral CP-even scalar states are obtained by orthogonal combinations of $\rho_1$ and $\rho_2$ given in Eq. \ref{eq:eq1}. A lighter $h^0$ and a heavier $H^0$ bosons are defined as $h^0 = \rho_1\sin\alpha-\rho_2\cos\alpha$ and $H^0 = \rho_1\cos\alpha-\rho_2\sin\alpha$. Accordingly, the SM Higgs boson would be 
		\begin{eqnarray}
			H^{\text{SM}}	&=& \rho_1\cos\beta+\rho_2\sin\beta\\
							&=&h^0\sba+H^0\cba
		\end{eqnarray}
        where the angle which rotates the CP-even Higgs states are defined as $\sba=\sin(\beta-\alpha)$, $\cba=\cos(\beta-\alpha)$. If $\sba=1$ is assumed, which is called the \emph{SM-alignment limit} \cite{Gunion:2002zf, Carena:2013ooa, Dev:2014yca, Bernon:2015qea}, an important feature shows up; the ratio of the couplings between $h^0$  ($H^0$) and the SM gauge bosons ($V=W^\pm/Z^0$) to the corresponding SM Higgs one will be $c_{VVh^0}/c_{VVH_{SM}}=\sba$ ($c_{VVH^0}/c_{VVH_{SM}}=\cba$), respectively. Therefore, the lighter CP-even Higgs boson ($h^0$) becomes indistinguishable from the Standard Model Higgs boson, and $H^0$ acts as gaugephobic ($\cba=0$) in this limit. In the literature, to explore the phenomenology of heavier CP-even boson ($H^0$), it is custom not to set $\sba$ to unity. That is also explored, and small deviations from unity are considered in the numerical calculation. The rest of the degree of freedom makes the prominent property of the model; two charged Higgs bosons and three neutral Higgs bosons \cite{Gunion:1989we}. In conclusion, the free parameters of the model are the masses of the neutral Higgs bosons $(m_{h/H^0/A^0})$ and the charged Higgs bosons ($m_{H^\pm}$), the ratio of the vacuum expectation values ($\tb=v_2/v_1$), the mixing angle between the CP-even neutral Higgs states ($\alpha$), and the soft breaking scale of the discrete symmetry $m_{12}$ \cite{Gunion:2002zf}.


\section{Calculating the cross section}
\label{sec:3}

    In this section, the analytical expressions, the vertices and the Feynman diagrams relevant to the scattering process $\theprocess$ are presented. Throughout this paper, the process is denoted as 
    \begin{eqnarray}
        e^+ (\mu)+e^- (\nu) \rightarrow H^+ (k_3)+ H^- (k_4) +Z^0(k_5),\nonumber
    \label{eq:eq3}
    \end{eqnarray}
    where $k_a$ $(a=3,4,5)$ are the four-momenta of the outgoing charged Higgs boson pair and $Z^0$ boson, respectively. Additionally, the positrons and the electrons are characterized by their spin polarization $\mu$ and $\nu$. Feynman diagrams which contribute to the process $\theprocess$ at the tree level are shown in Fig. \ref{fig:fig1} which is produced with the help of \texttt{FeynArts}. 2HDM Lagrangian and the corresponding vertices are calculated easily using \texttt{FeynRules} \cite{Alloul:2013bka, Degrande:2014vpa}. If there is a scalar particle in the model, there is a possibility of a scalar mediator between the incoming and outgoing states as it is seen in Fig. \ref{fig:fig1}. Then, these s-channel diagrams make a significant contribution, but they are almost negligible away from the mass pole of the mediator. In any case, the narrow-width-approximation for the scalars are employed, and the decay widths of the additional Higgs states are calculated with the help of \texttt{2HDMC} \cite{Eriksson:2009ws} for each scenario. The analysis revealed that the couplings $c_{H^+H^-h^0}$ and $c_{H^+H^-H^0}$ make the dominant contribution to the cross section. All the vertices involved in the scattering process are given in Tab. \ref{tab:tab1}.
    
    Another way to suppress the FCNCs in the Yukawa sector is to impose a natural condition which is to take the two Yukawa matrices to be aligned (Eq. (2.7) in \cite{Enomoto:2015wbn}), and the $Z_2$ symmetric types are assumed as the particular cases of this aligned model. There, the $\zeta_L$ is the factor defines the Yukawa coupling structure namely the Type-I through -IV. As follows, $\zeta_L=1/\tb$ in Type-I and Type-IV, and it is defined as $\zeta_L=-\tb$ in Type-II \cite{Enomoto:2015wbn} and -III. For that reason, the numerical results presented in this work hold for Type-III and -IV as well. Having defined all the couplings, the amplitude for each of the diagrams are constructed using \texttt{FeynArts}\cite{Kublbeck:1992mt, Hahn:2000kx}. Next, the simplification of the fermion chains, squaring the corresponding amplitudes, and the numerical analysis is accomplished using \texttt{FormCalc}\cite{Hahn:2006qw} routines.

        \begin{widetext}
        \onecolumngrid
        \begin{figure}[htbp]
        \centering
        \includegraphics[width=0.70\textwidth]{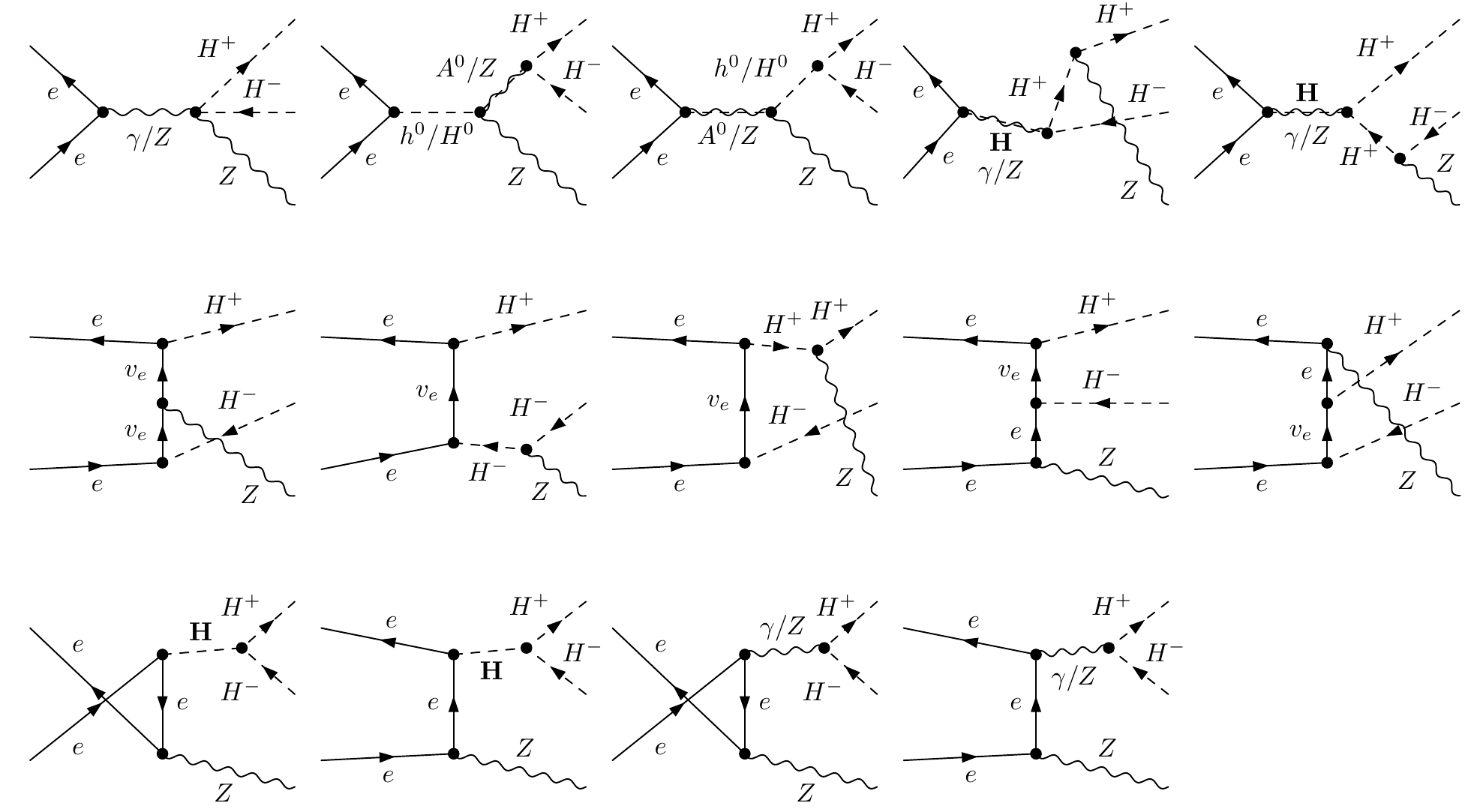}
        \caption{All the Feynman diagrams which contribute to the scattering process $\theprocess$ at the tree level. The dashed-wavelike line represents the nature of the propagating particle;  vector boson ($\gamma/Z$-boson) or scalar particle ($h^0/H^0/A^0$). The bold $\mathbf{H}$ represents any of $h^0/H^0/A^0$ bosons.}
         \label{fig:fig1}
        \end{figure}
        \end{widetext}

        \begin{table}[htp]
        \caption{The quartic and the trilinear couplings involved in the scattering process. The weak angle is defined as $s_w=\sin\theta_w, \;c_w=\cos\theta_w$. $s_{\alpha\beta}=\sin(\alpha+\beta)$, and $c_{\alpha\beta}=\cos(\alpha+\beta)$ are abbreviated. 
        }
        \begin{center}
        \begin{tabular}{ l | c l}
        \hline\hline    
        $c_{H^+H^-\gamma Z^0}$     & & $\frac{ie^2}{c_w s_w}(c_w^2-s_w^2)$                    \\
        $c_{H^+H^-Z^0Z^0}$         & & $\frac{ie^2}{2c_w^2 s_w^2}(c_w^2-s_w^2)^2$                       \\
        $c_{Z^0Z^0[h^0,H^0]}$      & & $\frac{i e m_w}{2 c_w^2 s_w}[\sba,\cba]$                  \\
        $c_{H^+H^-\gamma}$         & & $i e$                                              \\
        $c_{H^+H^-Z^0}$            & & $\frac{ie}{2c_w s_w}(c_w^2-s_w^2)$                                \\
        $c_{H^+H^-h^0}$            & & $\frac{i}{v} \left((\mhl^2 - 2 \mhp^2)s_{\beta\alpha} -  (2 \mhl^2 - \frac{2m_{12}^2}{s_\beta c_\beta})\frac{c_{\alpha\beta}}{s_{2\beta}}\right)$ \\
         $c_{H^+H^-H^0}$             & & $\frac{i}{v} \left((\mhh^2 - 2 \mhp^2)c_{\beta\alpha} -  (2 \mhh^2 - \frac{2m_{12}^2}{s_\beta c_\beta})\frac{s_{\alpha\beta}}{s_{2\beta}}\right)$ \\
       
        $c_{e \bar{e} [h^0,H^0]}$        & & $-\frac{i m_e }{v} \left[(\sba+\cba\zeta_L),(-\cba+\sba\zeta_L) \right]$    \\
        $c_{\nu_e \bar{e} H^+}$        & & $\frac{i \sqrt{2} m_e }{v}\zeta_L $                   \\
        \hline\hline    
        \end{tabular}
        \end{center}
        \label{tab:tab1}
        \end{table}
	After squaring the amplitude, a summation over the polarization vectors of the final states and averaging over the helicities of the initial states is performed. Finally, the total unpolarized cross section is defined as an integral over a 4-fold differential cross section \cite{Kublbeck:1990xc}
   \begin{eqnarray}
   \sigma = \frac{1}{4} \sum_{\mu,\nu} && \int_{m_{Z^0}}^{(k_5^0)_\text{max}} dk_5^0\int_{(k_3^0)_{min}}^{(k_3^0)_\text{max}}dk_3^0 \int_{-1}^{+1} d\cos\theta \nonumber \\
            &&\times \int_0^{2\pi} d\eta\frac{d^4\sigma_{\mu\nu}}{dk_5^0 dk_3^0 d\cos\theta d\eta}
    \end{eqnarray}
    where the energy of the $k_3$ and $k_5$ are defined below and $a=\sqrt{s}-k_5^0$.
    \begin{eqnarray}
    (k_3^0)_\text{min,max}&=&\frac{1}{2}\left[ a\pm |\vec{k_5}| \sqrt{1-\frac{4m_{H^\pm}^2}{a^2-|\vec{k_5}|^2 }} \right] \\
    (k_5^0)_\text{max}&=&\frac{s-2m_{H^\pm}^2+m_{Z^0}^2}{2\sqrt{s}}.
    \end{eqnarray}

    One other option at the LC is to polarize the incoming beams which could maximize the physics potential, both in the performance of precision tests and in revealing the properties of the new physics beyond the SM. In this study, we also explored the dependence of the cross section on the polarization of the incoming electron and positron beams. 
	The polarization of the incoming beam is significant especially to enhance some of the helicity channels. In annihilation diagrams, typically in s-channel, the helicities of the incoming beams are coupled to each other. For that reason, the helicities of the incoming particles in the SM need to be opposite from one another to recombine into the vector boson mediator, the $Z^0$ boson or the photon. In exchange diagrams, the helicities of the incoming beams are directly coupled to the helicities of the final states. In this case, all helicity configurations for the beams are in principle possible.

    The expression for the cross section for an arbitrary degree of longitudinal beam polarization is defined as:
    \begin{eqnarray}
    \sigma(P_{e^+}, P_{e^-} )= \frac{1}{4}[
        &+& (1-P_{e^-})(1+P_{e^+})\sigma_{LR} \nonumber\\
        &+& (1+P_{e^-})(1-P_{e^+})\sigma_{RL} ]
    \label{eq:eqpolcross}
    \end{eqnarray}
    where $\sigma_{LR}$ denotes the cross section for 100\% left-handed positron and 100\% right-handed electron polarization. $P_{e^-}$ and $P_{e^+}$ denote the percentage of the electron and positron beam polarization, respectively.  The $\sigma_{LL}$ and $\sigma_{RR}$ configurations are omitted due to the negligible contributions in Eq. \ref{eq:eqpolcross}.
	In this study, we examined various polarization configurations in an LC and presented in the next chapters. Two of these configurations are inspired from the ILC which in pol-1 (pol-2) is defined as right-handed positron with 30\% (60\%) polarization \cite{Fujii:2015jha, CLIC:2016zwp} and an electron with 80\% left-handed polarization on both cases, respectively.
    
    \begin{table}[htp]
    \caption{Polarization configuration of the incoming beams used in this study are denoted by left (-) and right (+).}
    \begin{center}
    \begin{tabular}{l | l c}
    \hline
    pol-1 & $P_{e^+}, P_{e^-}$ & (+0.3,\;-0.8)\\
    pol-2 & $P_{e^+}, P_{e^-}$ & (+0.6,\;-0.8)\\
    \hline    
    \end{tabular}
    \end{center}
    \label{tab:tab2}
    \end{table}%


\section{Constraints from theory and experiments, benchmark points}
\label{sec:4}

    There are two sets of constraints in 2HDM; one is the theoretical constraints which come from the theory itself, the other one is the results coming from the measurements carried out in the past and current experiments. These constraints need to be applied to the free parameters defined in the previous section.

\subsection{Theoretical Constraints}

    \begin{itemize}
    \item Perturbative Unitarity: This requirement comes from the fact that the scattering amplitudes need to be flat at the asymptotically high energies. Due to the additional Higgs states in 2HDM, we need to make sure that Higgs-Higgs and Higgs-Vector boson scattering cross sections are bounded by $16\pi$ \cite{ginzburg2005tree}.  
    \item Perturbativity: The theory needs to be inside the perturbative region. The perturbative region is defined as where all the quartic couplings in the theory are small, and we take them to be $|\lambda_i| \leq 4\pi$.
    \item Vacuum Stability: The scalar potential defined in Eq. \ref{eq:eq2} needs to be positive in any direction of the field space, even at the asymptotically large values \cite{el2007consistency, sher1989electroweak, deshpande1978pattern, Nie:1998yn, ElKaffas:2006gdt}. Stability constraint translated into the following conditions: 
    \begin{eqnarray}
        \lambda_1>0, \;\;\; \lambda_2>0, \nonumber \\
        \lambda_3+\sqrt{\lambda_1\lambda_2}>0, \nonumber \\
        \sqrt{\lambda_1\lambda_2}+\lambda_3+min(0,\lambda_4-|\lambda_5|)>0 \nonumber
    \end{eqnarray}
    \end{itemize}

\subsection{Experimental Constraints}

    \begin{itemize}
    \item ATLAS and CMS experiments at the LHC have reported the so long hunted resonance with a mass of $125.4\pm0.4\gev$ \cite{chatrchyan2012observation, Aad:2012tfa, Khachatryan:2014ira, Chatrchyan:2012xdj, Khachatryan:2014jba, Aad:2015gba}. We know that the announced peak must have a nature of a CP-even, and for that reason 125.4 GeV peak needs to correspond to the one of the CP-even $h^0$ or $H^0$ states in 2HDM. In the numerical calculation, we considered both of these possibilities and investigated the implications on the cross section of $\theprocess$. If we assume that $h^0$ is the discovered particle at the LHC, that puts a restraint on the couplings. Therefore, $s_{\beta-\alpha}$ is pushed to unity. In the opposite case, where the $H^0$ is assumed that it is the discovered one then $\cba$ is set to unity. However, we let that factor to deviate from unity just for the sake of phenomenological curiosity.
 
   \item 2HDM needs to be compatible with the full set of electroweak precision observables which are measured in the previous experiments \cite{ALEPH:2005ab}. There are parameters S, T, and U which are called oblique parameters \cite{Peskin:1990zt, Peskin:1991sw}, and they merely represent the radiative corrections to the two-point correlation functions of the electroweak gauge bosons. These parameters are sensitive to any new physics contribution, and they take the value of the top-quark and Higgs masses so that they are set to vanish for a reference point in the SM ($S=T=U=0$). According to that, a sizable deviation from zero would be an indicator of the existence of new physics. These oblique parameters have been calculated by \cite{Haber:2010bw, gunion2003cp, Grimus:2008nb, Polonsky:2000rs} for the scenarios presented in the next section with the help of \texttt{2HDMC}, and in all cases the oblique parameters are much less than $10^{-2}$. 
 
   \item The LEP experiment excluded the charged Higgs boson with a mass below $80\gev$ (Type II scenario) or $72.5\gev$ \cite{Searches:2001ac} (Type I scenario, for pseudo-scalar masses above $12\gev$) at the 95\% CL. If it is assumed that $\mathcal{B}(H^+\rightarrow \tau^+\nu)=1$, then charged Higgs mass bound increased to $94\;GeV$ for all $\tan\beta$ values. \cite{Abbiendi:2013hk}. The Tevatron experiments D0 \cite{Abazov:2008rn, Abazov:2009ae, Abazov:2009wy} and CDF \cite{Aaltonen:2009ke} excluded the charged Higgs mass in the range of $80\gev<m_{H^\pm}<155\gev$ at the 95\% CL assuming $\mathcal{B}(H^+ \rightarrow c\bar{s})$. The search on charged Higgs is also carried out at the LHC in the decay of top quark \cite{Aad:2012tj, Chatrchyan:2012vca}, and upper limits are set on the $\mathcal{B}(t\rightarrow H^+ b)$ and $\mathcal{B}(H^+\rightarrow \tau\nu)$. More recent results are given in \cite{Moretti:2016qcc} and the references therein.
  
  \item It is known that charged scalar states in 2HDM affects the flavor physics, particularly $B_s \rightarrow X_s\gamma$ or $B_s \rightarrow \mu\mu$. In general, the flavor observables in these models are sensitive to the $m_{H^\pm}$ and $\tan\beta$. According to Refs. \cite{ElKaffas:2007rq, WahabElKaffas:2007xd}, $\bar{B}-B$ mixing disfavors $\tan\beta<0.5$, and also lets the couplings of the Higgs to heavy quarks to be in the perturbative region $\tb<60$ \cite{Akeroyd:1995hg, Barger:1989fj}. More discussion is carried out in Ref. \cite{Searches:2001ac} and the references therein.
	\end{itemize}

\subsection{Benchmark Points}

    Recognizing all these experimental and theoretical constraints,  the following scenarios and benchmark points are chosen. These points are preferred by aiming at a broader survey of the region of phenomenologically interested. As it is stated in the experimental constraints, we employed the \texttt{2HDMC} \cite{Eriksson:2009ws} to check whether the theoretical constraints for each benchmark point are fulfilled.

    \begin{itemize}
    \item \textbf{\emph{Non-alignment scenario}} : This benchmark point is taken from Ref. \cite{Haber:2015pua} where the particle with the mass of 125 GeV is interpreted as the lightest CP-even Higgs boson ($h^0$) with SM-like couplings. In the so-called alignment limit where $|\cba| \rightarrow 0$ the $c_{hVV}$ coupling approaches the corresponding SM value, and in which case the heavier CP-even Higgs boson $H^0$ could not decay into $W^+W^-/ZZ$. The alignment limit is also endorsed by the additional Higgs boson searches at the Large Hadron Collider. However, to allow some interesting phenomenology for the heavier CP even state ($H^0$), this benchmark is defined with a non-alignment ($\cba \neq 0$) as allowed by the present constraints. With this motivation, the mass of the other Higgs states $A^0$ and $H^\pm$ are taken as degenerate and they are let to decouple $m_{h^0} =125<m_{H^0}<m_{A^0}=m_{H^\pm}$. 
        Detailed analysis is carried out in \cite{Haber:2015pua} and the charged Higgs mass is given in the Hybrid base as 
        \begin{equation}
        \mhp^2=\maa^2=\mhh^2 \sba^2 + m_{h^0}^2 \cba^2-Z_5 v^2 \label{eq:eq10}
        \end{equation}
         where $Z_5$ is the quartic coupling parameter, and it is set to $-2$ along with $Z_4$. In this scenario, $\tb$ and and $\mhh$ are taken as a free parameters and their ranges, as well as the other parameters, are given in Tab. \ref{tab:bp_nonalignment}.
        \begin{widetext}
         \onecolumngrid       
        \begin{table}[htp]
        \caption{Benchmark points for the non-alignment scenario, all masses are given in GeV.}
        \centering
        \begin{tabular}{ c | ccc l cccc}       
        \hline         
        Bench. & Yuk. T.& $m_{h^0}$ & $m_{H^0}$ & $\cba$  & $Z_4$    & $Z_5$ & $Z_7$ &  $\tb$     \\\hline \hline        
        1    &    Type-I   & 125 & $(150\ldots600)$ & 0.1                                            & -2& -2   & 0    & (1...50)    \\\hline
        2    &  Type-II  & 125 & $(150\ldots600)$ & $0.01\left( \frac{150\gev}{\mhh} \right)^2$    & -2& -2   & 0    & (1...50)    \\\hline\hline
        \end{tabular}
        \label{tab:bp_nonalignment}
        \end{table}
        \end{widetext}
    \item \textbf{\emph{Low-$m_H$ mass scenario}} : As it is known, there are two CP-even states ($h^0/H^0$) in 2HDM. In the previous scenario, it is assumed that the discovered scalar particle at the LHC in 2012 is the lighter CP-even Higgs ($h^0$) state. However, there is one other possibility that it could be the heavier CP-even Higgs ($H^0$) state. In that case, the coupling of the heavier CP-even Higgs to gauge bosons ($c_{HVV}$) will be scaled by a factor of $\cba$ instead of $\sba$, and as in the previous case that also forces us to $\cba\approx\pm1$. Due to the direct search limits, couplings of $h^0$ to vector bosons need to be suppressed heavily and even close to zero ($\sba \rightarrow 0$). 
        This scenario is analyzed in detail in Ref. \cite{Haber:2015pua} where the region $90 < m_h < 120$ GeV is not rejected by the LHC constraints (from $h^0 \rightarrow bb,\tau\tau$), which puts an upper limit on $\tb$.
        According to the authors, it is also possible to set exact alignment ($\cba=1$) with either Type-I or Type-II Yukawa couplings. Therefore, choosing a non-aligned value for $\cba$ doesn't make much impact on the production of the cross section, but more on that is delivered in the results. In Tab. \ref{tab:bp_lowm_h_mass}, the free parameters are depicted in the Hybrid base where $Z_4=Z_5$ is taken so that in particular the $T$ oblique parameter can not receive sizable contributions (Eq. (76) in Ref. \cite{Haber:2015pua}).

        \begin{table}[htp]
        \caption{Benchmark points for the low-$m_H$ mass scenario, all masses are given in GeV.} 
        \centering
        \begin{tabular}{ c | ccc l cccc}
        \hline         
        Bench. & Yuk. T.& $m_{h^0}$ & $m_{H^0}$ & $\cba$  & $Z_4$ & $Z_5$ & $Z_7$ &  $\tb$     \\\hline \hline        
        1    & Type-I   & $(65\ldots 120)$ & 125  & 1.0    & -5    & -5     & 0    & 1.50    \\\hline
        2    & Type-II  & $(65\ldots 120)$ & 125  & 1.0     & -5    & -5     & 0    & 1.50    \\\hline\hline
        \end{tabular}
        \label{tab:bp_lowm_h_mass}
        \end{table}

    \item \textbf{\emph{Favored region in light of the recent experimental constraints}}: Taking into account all the recent updates particularly coming from the flavor physics, as the last scenario, we explored the region inspired by the results presented in Ref. \cite{Enomoto:2015wbn}. Since the charged Higgs boson can contribute to flavor observables via the charged currents, flavor observables are quite significant. Motivated by the Ref. \cite{Enomoto:2015wbn}, the same masses for all the Higgs bosons are set $m_{H^0}= m_{A^0} = m_{H^\pm}$, and that also satisfies the theoretical constraints. The $\sba$ is set approaching the unity which guarantees that all the light Higgs self-couplings are close to the SM ones. In that case, the heavier CP-even Higgs boson ($H^0$) can not decay into $W^+W^-$ and $ZZ$ pairs. According to the authors, in Type-I, $\tb<1$ is strongly constrained by $\mathcal{B}(B_s^0 \rightarrow \mu^+\mu^-)$ and the mass difference of the scalars. However, the mass of the scalars is not constrained on the large $\tb$ range compared to the Type-II. The cross section in Type-II, as in the Type-I, is calculated by setting the same masses for all the extra scalars. $\sba$ is taken to unity, and due to the dominant constraints from $\mathcal{B}(B \rightarrow \tau\nu)$ and $\mathcal{B}(B_q^0 \rightarrow \mu^+\mu^-)$ high $\tb$ are excluded. The parameter region is defined in Tab. \ref{tab:flavor_inspired}, and more detailed discussion is given in \cite{Enomoto:2015wbn} (particularly Fig. 3 and the references therein). In this basis, one other parameter required for the calculation is the $m_{12}^2$. The region for the $m_{12}^2$, where the theoretical constraints are fulfilled, is calculated with the help of \texttt{2HDMC}.

    \begin{table}[htp]
    \caption{Benchmark points for the favored region in light of the recent experimental constraints in flavour physics, all masses are given in GeV.} 
    \centering
    \begin{tabular}{ c | ccccc }       
    \hline
    Bench.        & Yuk. T.    & $m_{h^0}$    &$m_{(H^0/A^0/H^\pm)}$    &  $\sba$     &  $\tb$     \\\hline \hline        
    1             & Type-I     & 125         & (150..1000)             & 1.0        &     (1..50)    \\\hline
    2             & Type-II    & 125         & (500..1000)             & 1.0        &     (1..40)    \\\hline\hline
    \end{tabular}
    \label{tab:flavor_inspired}
    \end{table}

    \end{itemize}

\section{Numerical results and discussion}
\label{sec:5}
    In this section, the numerical results for the charged Higgs pair production associated with the $Z^0$-boson in an $e^+e^-$-collider are presented and discussed for each scenario. The current parameters of the SM are taken from Ref. \cite{Eidelman:2004wy} where $m_e=0.51099892\text{ MeV}$, $m_Z=91.1876\gev$, $s_w=0.222897$, and $\alpha=1/137.035999$ are set. Taking into account all these constraints, the numerical analysis is carried out for unpolarized and polarized incoming beams.

\subsection{Non-Alignment Scenario}
    
        \begin{figure}[htbp]
        \centering
        \includegraphics[width=\factor\textwidth]{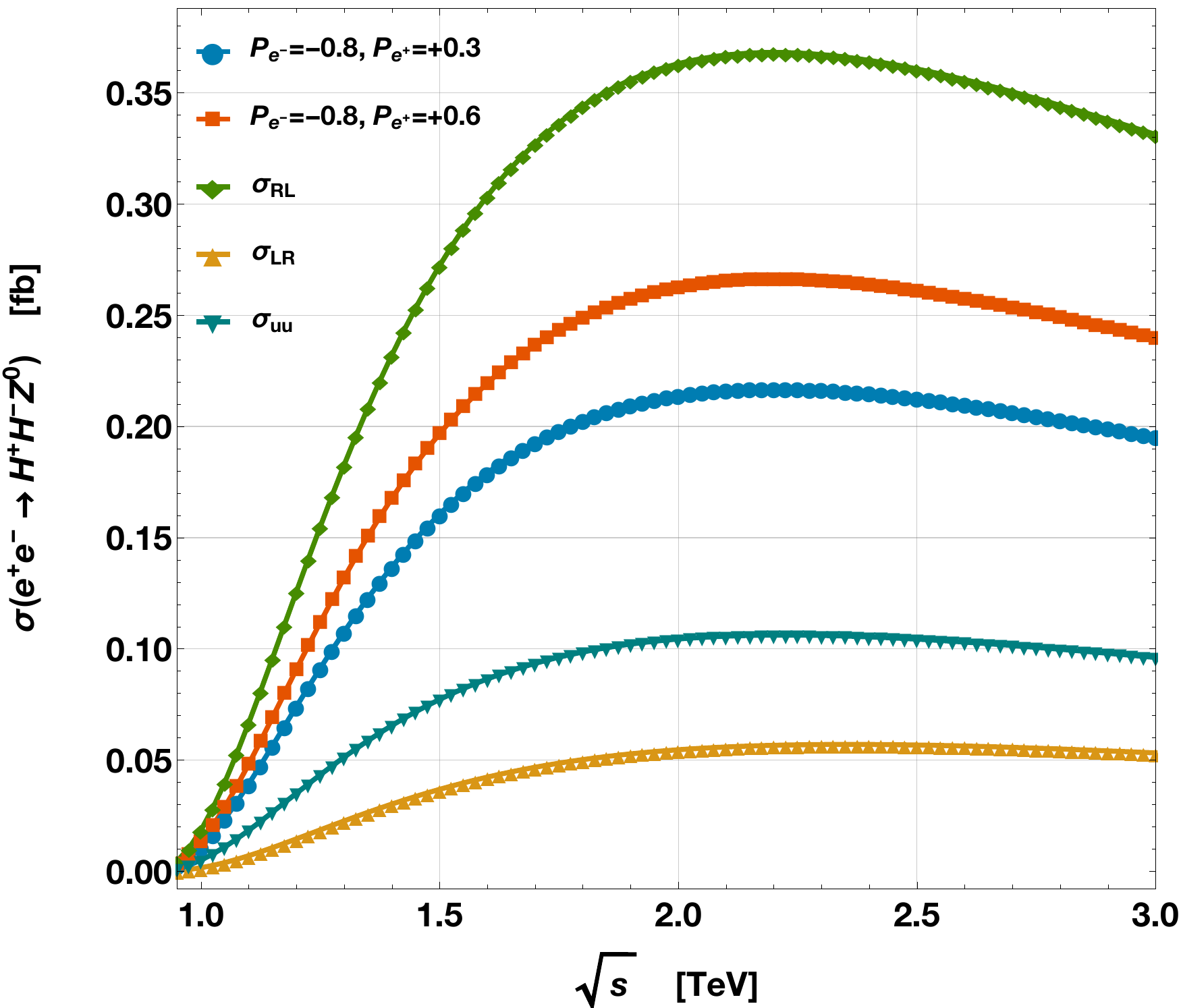}
        \includegraphics[width=\factor\textwidth]{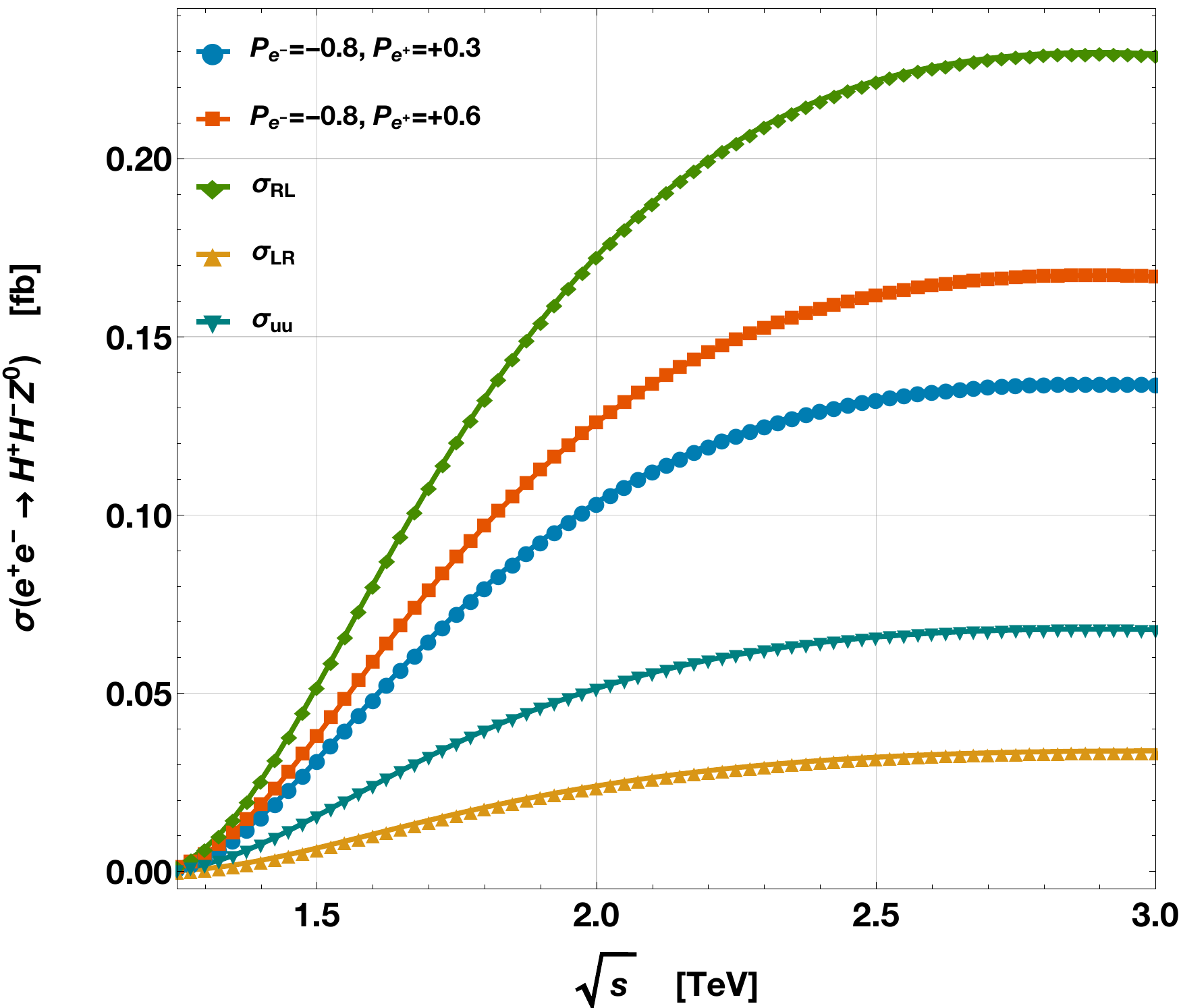}
        \caption{        \label{fig:fig2}
        Comparison among various polarization cases for two benchmark points in the non-alignment scenario. 
        The polarization configurations are depicted in the figure.
        (left): Where $\cba=0.1$, $\tb=45$, $\mhh =  200\gev$, and Type-I is set. $\sigma_{LR}$, $\sigma_{RL}$ and $\sigma_{UU}$ represent the possible helicity configuration of the incoming positron and electron beams.
        (right): the same caption with the left figure, but $\cba=0.01\left( \frac{150}{m_{H^0}}\right)^2$, $\mhh= 425\gev$, and Type-II is assumed.
        }
        \end{figure}

    The cross section of the process ($\theprocess$) is computed for the parameters presented in the non-alignment scenario \cite{Haber:2015pua}. In Fig. \ref{fig:fig2}, we plotted the cross section as a function of COM energy in $1<\sqrt{s}<3 \tev$ range and for two distinct charged Higgs masses ($m_{H^\pm}$). We also explored the polarization configurations of the incoming beams. As it is seen in Fig. \ref{fig:fig2}, the cross section $\sigma_{RL}$, where it represents the totally right-handed polarized $e^+$-beam and totally left-handed polarized $e^-$-beam, is always greater among other polarization cases. In Fig. \ref{fig:fig2} (left), the cross section for polarized incoming beams has a value of $\sigma_{RL}\approx 0.368 \fb$ around $\sqrt{s}= 2.2\tev$, whereas in the opposite helicity configuration the cross section is $\sigma_{LR}\approx 0.057 \fb$. The neutral Higgs mass is set to $m_{H^0}=200\gev$, and the Yukawa couplings are in Type-I configuration. The total unpolarized cross section ($\sigma_{UU}$) is calculated, and it is around $0.106 \fb$. Since the polarization of the colliding beams lets us study and tune various couplings of the model, we also produced the distribution of the cross section for two possible polarization cases in the LC. The cross section have a value of $\sigma_{\text{pol-1}}\approx 0.217\fb$, and in the other polarization configuration where the positron is polarized right-handed by 60\% the cross section gets up to $\sigma_{\text{pol-2}}\approx 0.266 \fb$ for the Type-I in Fig. \ref{fig:fig2} (left). A large separation between $\mhp$ and $\maa$ is strongly constrained by the vacuum stability and perturbativity, for that reason, the calculation is carried in the limit of $\maa \sim \mhp$ in the non-alignment scenario. Therefore, any value of $\maa$ is allowed by the EW precision tests.

    In Fig. \ref{fig:fig2} (right), the production rate of $\products$ is plotted for various polarization configurations, and the Type-II is set. The unpolarized cross section is $\sigma_{UU}=0.068\fb$ at $\sqrt{s}=2.9\tev$. However, the polarized cross sections are $\sigma_{\text{pol-1}}\approx 0.137\fb$ and $\sigma_{\text{pol-2}}\approx 0.167\fb$.

        \begin{figure}[htbp]
        \centering
        \includegraphics[width=\factor\textwidth]{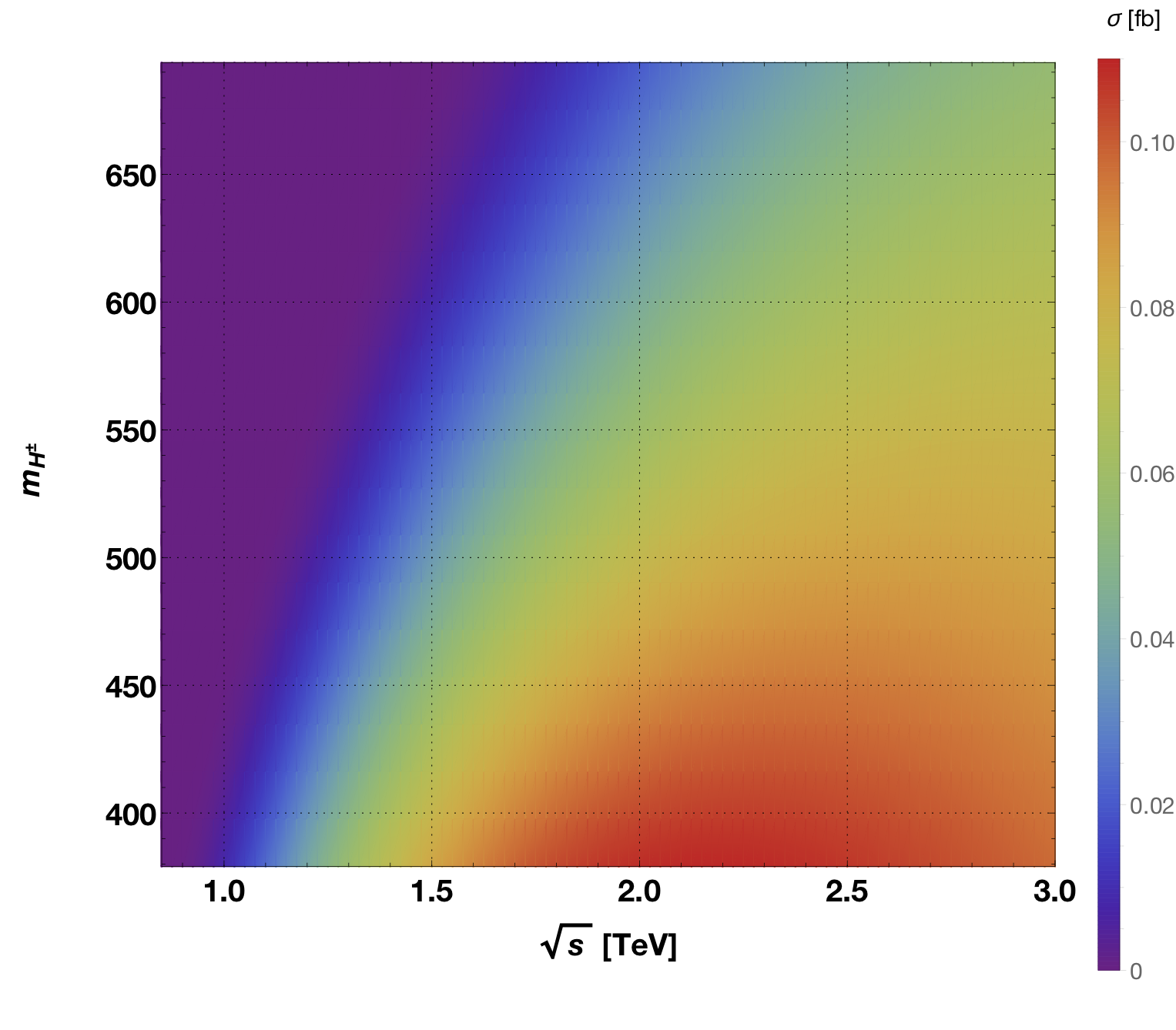}
        \includegraphics[width=\factor\textwidth]{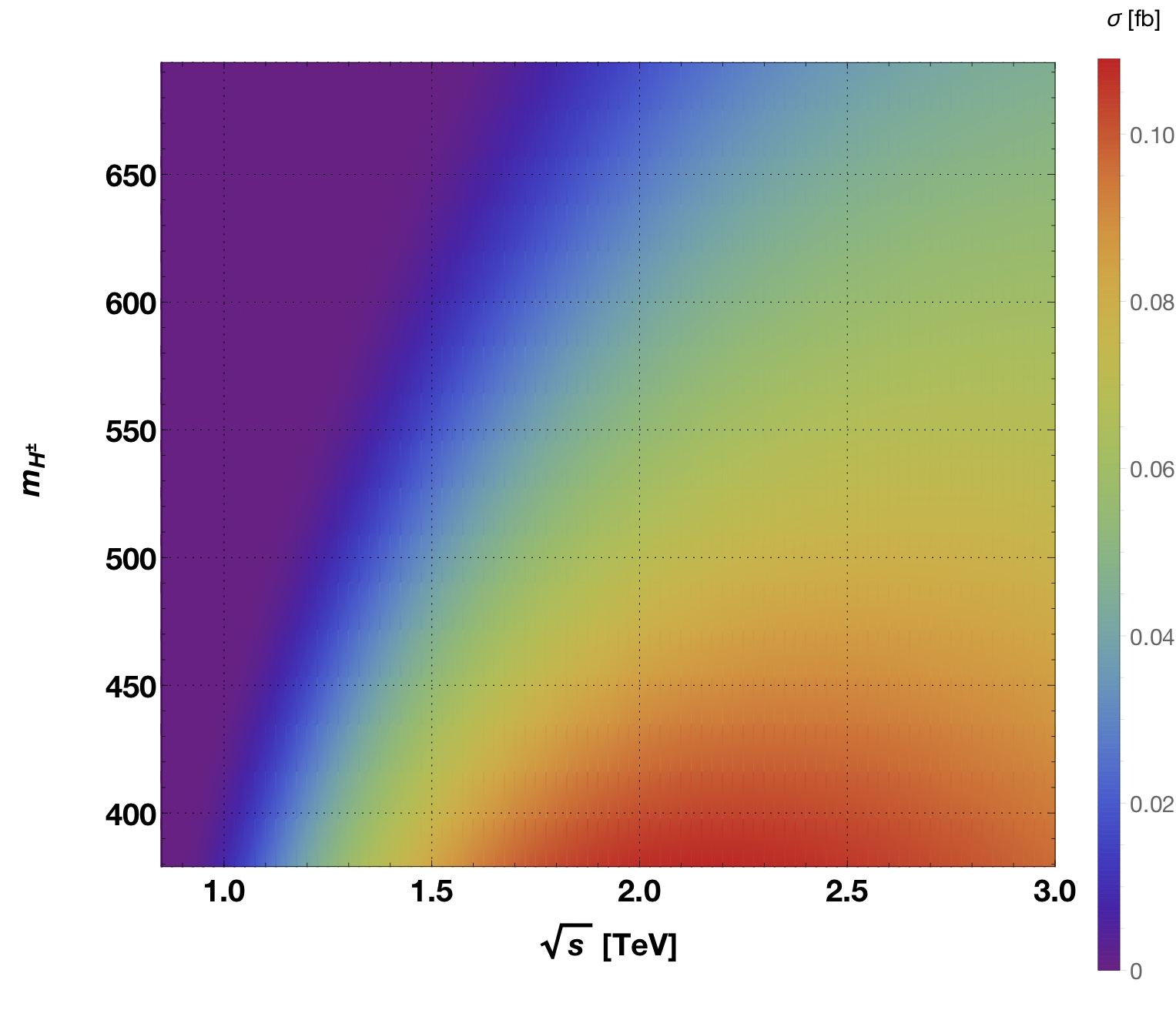}
        \caption{        \label{fig:fig3}
        The cross section is given for the unpolarized incoming beams as a function of the $\mhp$ and $\sqrt{s}$ in the non-alignment scenario. 
        The scan is done for $\tb=45$ and $\mhp$ is calculated by varying the $\mhh$.
        (left): $\cba=0.1$ and Type-I couplings,
        (right): $\cba=0.01\left( \frac{150}{m_{H^0}}\right)^2$ and Type-II couplings are considered.
        }
        \end{figure}

    Two-dimensional analysis of the cross section of the process is drawn in Fig. \ref{fig:fig3} as a function of the COM energy and the charged Higgs mass ($\mhp$). In Fig. \ref{fig:fig3} (left), the non-alignment scenario with Type-I Yukawa couplings are set, on the (right) the same analysis with Type-II couplings are displayed. The charged Higgs mass is computed by Eq. \ref{eq:eq10}, and we scanned in the allowed range $379 < \mhp < 690 \gev$ for this scenario. According to the couplings given in Tab. \ref{tab:tab1}, the $\mhp$ dependence appears only in the CP-even to charged Higgs pair couplings. Indeed, in the non-alignment scenario, $c_{H^+H^-h^0}$ is the dominant coupling affecting the cross section. Choosing the Type-I over the Type-II effects only the $c_{e^+\nu_e H^+}$ and the $c_{e\bar{e}[h^0,H^0]}$ couplings. Since each of them are a function of $t_\beta$, they will be boosted at high $t_\beta$ values in Type-I. Comparing the two plots in Fig. \ref{fig:fig3}, the cross section in Type-I (left) and Type-II (right) gets larger at low $\mhp$ values, and it declines at high $\mhp$. On the other hand, the fall is a little bit faster in Type-II due to the $1/t_\beta$ dependence of the couplings.

        \begin{figure}[htbp]
        \centering
        \includegraphics[width=\factor\textwidth]{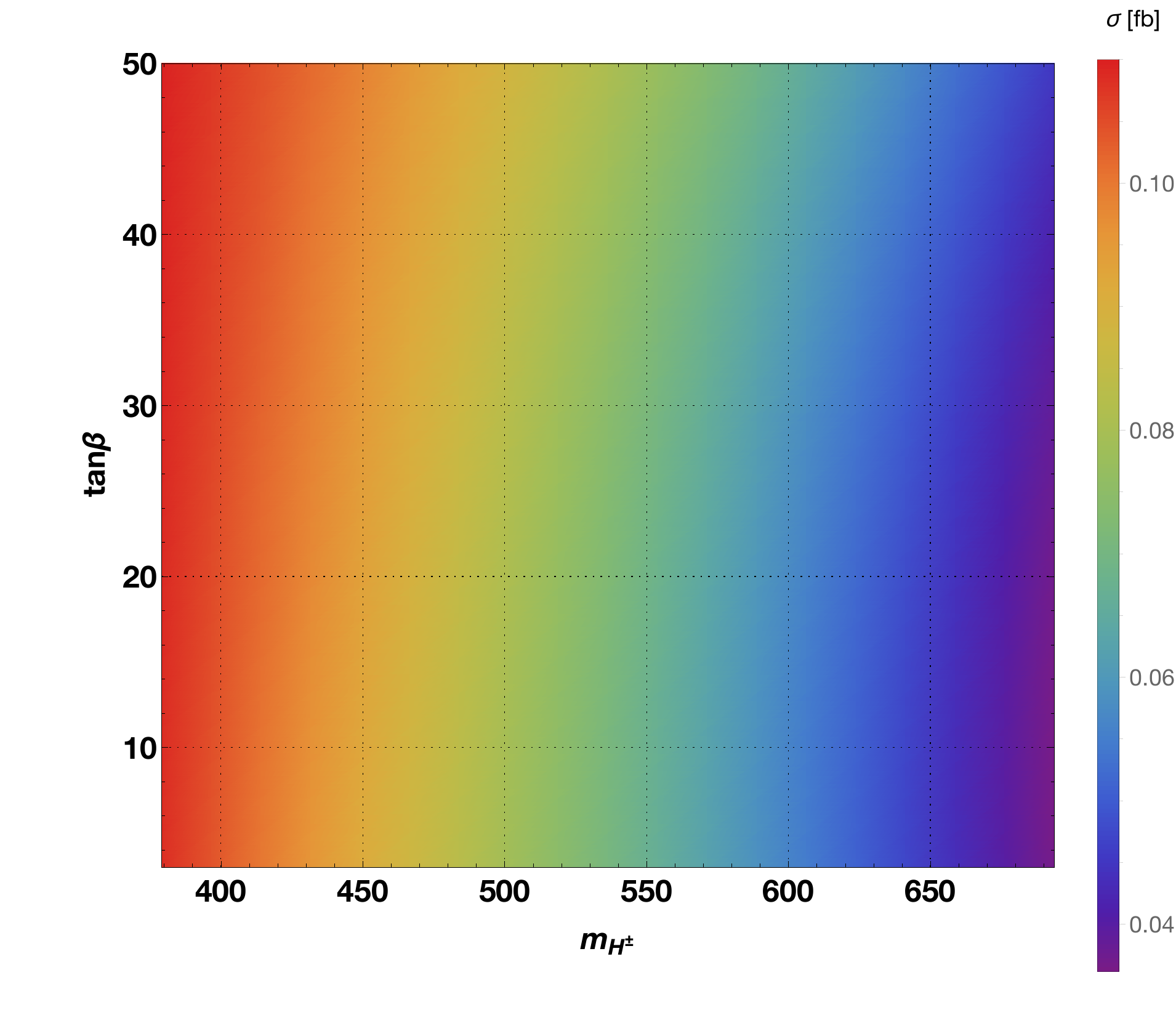}
        \includegraphics[width=\factor\textwidth]{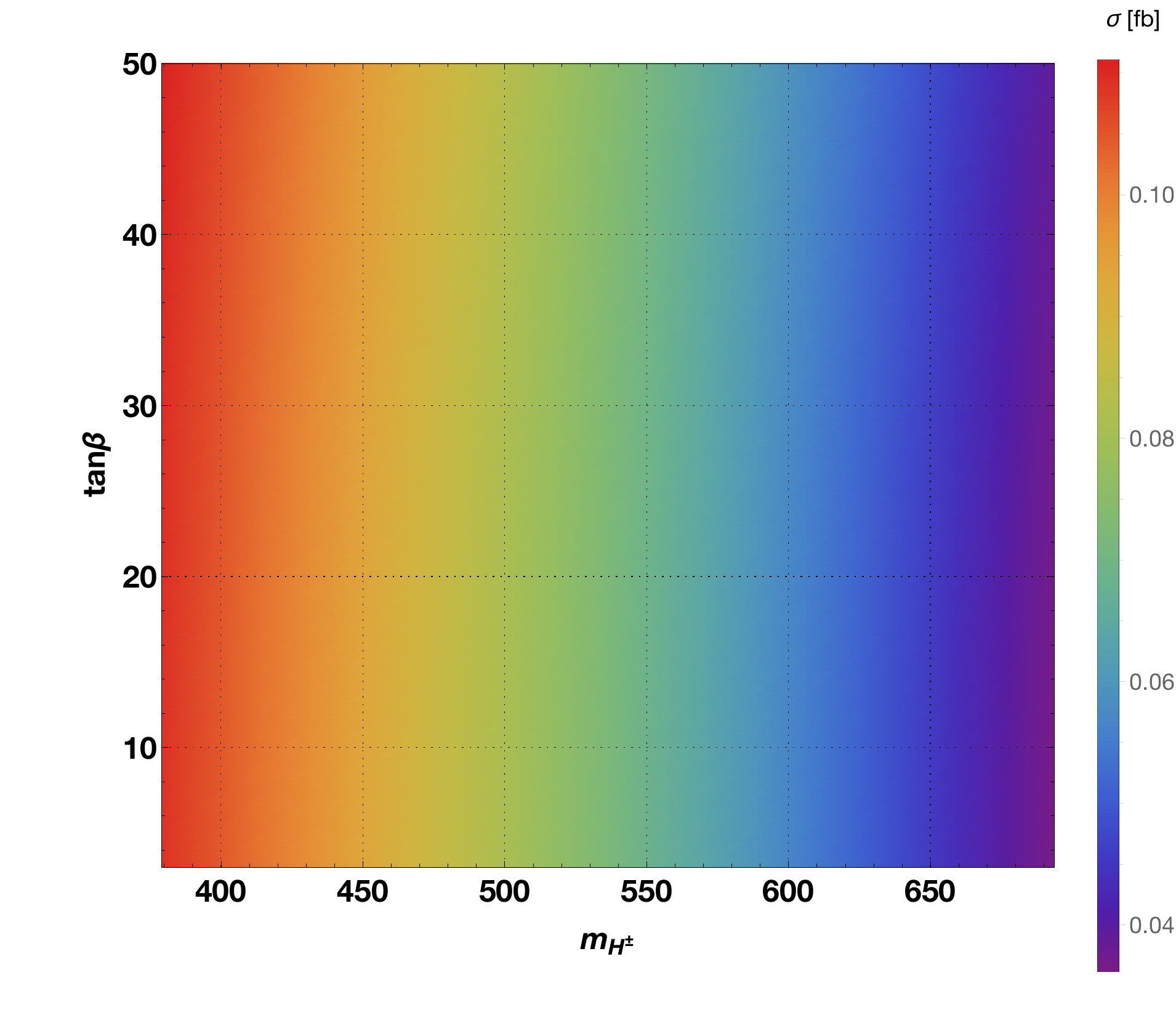}
        \caption{\label{fig:fig4}
        The unpolarized cross section at $\sqrt{s}=2.5\tev$ as a function of the $\mhp$ and $\tb$ in the non-alignment scenario;
        (left): Type-I Yukawa couplings with $\cba=0.1$,
        (right): Type-II Yukawa couplings with $\cba=0.01\left( \frac{150}{m_{H}}\right)^2$ are set.
        }
        \end{figure}

    The last analysis for this scenario is drawn in Fig. \ref{fig:fig4} where the unpolarized cross section at $\sqrt{s}=2.5\tev$ as a function of $\mhp$ and $\tb$ are given. If Type-I Yukawa couplings are chosen, the distribution of the cross section in Fig. \ref{fig:fig4} (left) shows a small dependence on the $\tb$ at high $\tb$ values compared to the Type-II Fig. \ref{fig:fig4} (right). On the (right), the cross section does not change much with $\tb$, but $\mhp$. The situation in Type-I happens due to the large values of the $\tb$ in $c_{e\bar{e}[h^0,H^0]}$ and $c_{\nu_e \bar{e}H^+}$ couplings (through $\zeta_L$) and $\mhp$ dependence of the $c_{H^+ H^-[h^0,H^0]}$ couplings. In each case, the cross section gets up to $\sigma_{UU} \approx 0.113 \fb$ at low $\mhp$ values. However, at smaller COM energies the distinction between them completely disappears even at high $\tb$ values.

\subsection{Low-$m_H$ Scenario}

\begin{figure}[htbp]
    \centering
    \includegraphics[width=\factor\textwidth]{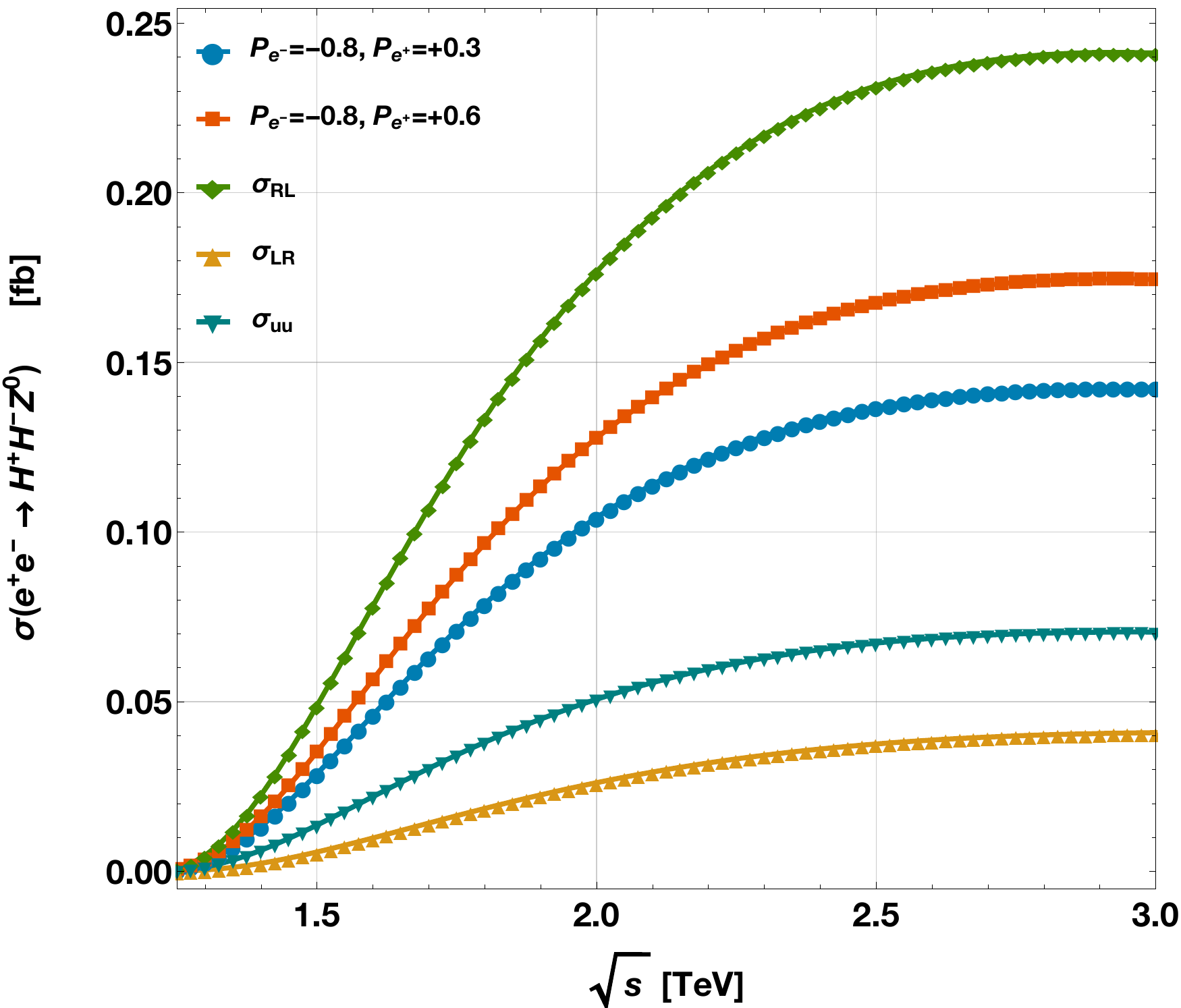}
    \includegraphics[width=\factor\textwidth]{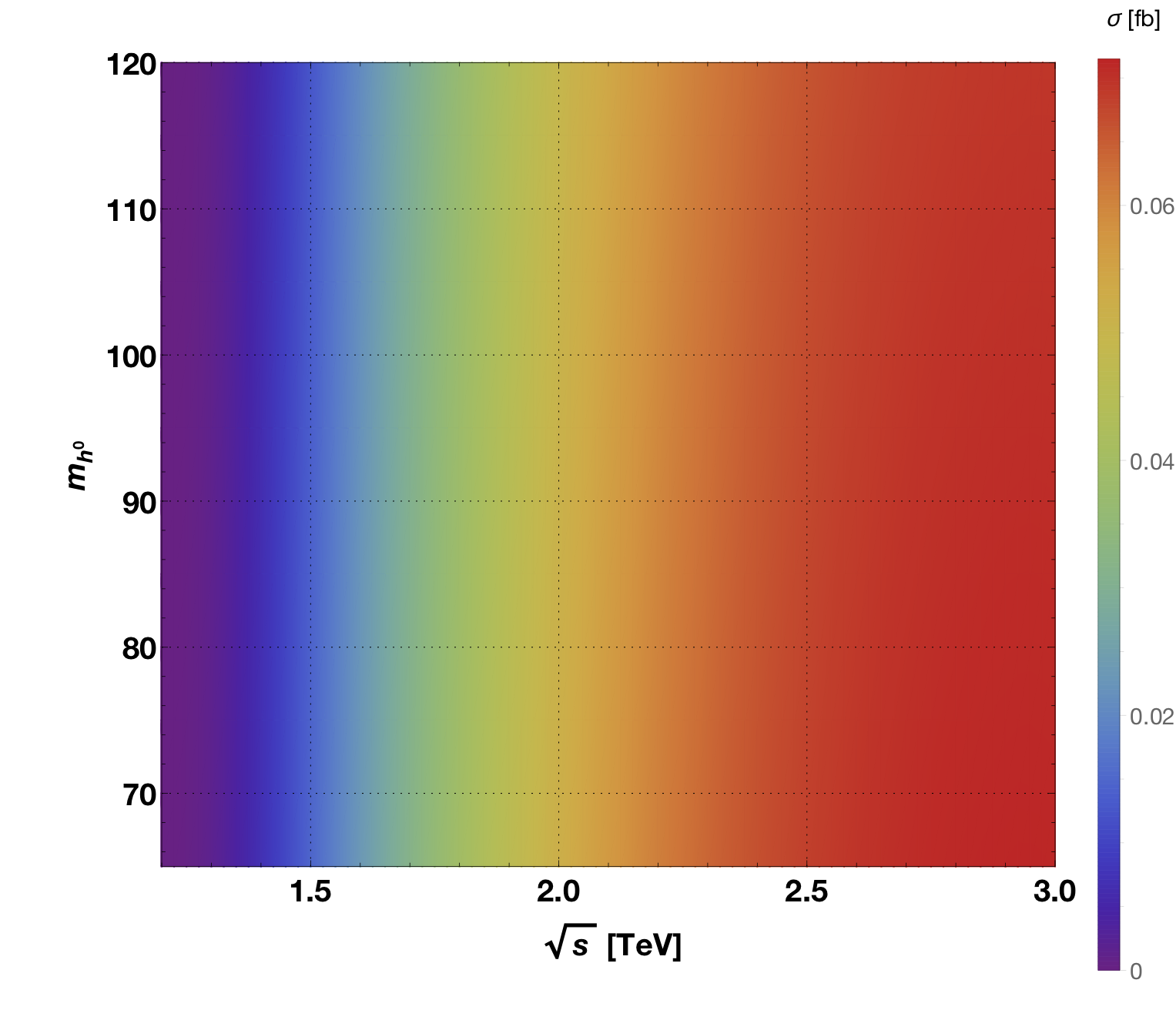}
    \caption{    \label{fig:fig5}
    The cross section distributions in the low-$m_H$ scenario, where $\mhl=95\gev$.
    (left): The process as a function of $\sqrt{s}$ is plotted for various polarizations depicted in the plot,
    (right): the cross section as a a function of light Higgs mass ($m_{h^0}$) and $\sqrt{s}$,
    }
    \end{figure}

    In this scenario, CP-even states are flipped where $H^0$ state is considered as the one who behaves like the SM Higgs boson. The computation is carried out for $\tb=1.5$, and quartic parameters are taken as $Z_4=Z_5=-5$ so that the CP-even Higgs states are decoupled from $A^0$ and $H^\pm$. According to Eq. (76) in Ref. \cite{Haber:2015pua}, that configuration also agrees with the vacuum stability and the perturbativity constraints. The cross section as a function of $\sqrt{s}$ is given in Fig. \ref{fig:fig5} (left). The unpolarized cross section gets up to $0.071 \fb$ at $\sqrt{s}=3.0\tev$ and then falls down, while the polarized cross sections are $\sigma_{LR}\approx 0.041 \fb$ and $\sigma_{RL}\approx 0.241 \fb$. The same with the non-alignment scenario, the cross section is maximized for configuration of a right-handed polarized $e^+$ and left-handed polarized $e^-$ beams. The unpolarized cross section as a function of the light Higgs mass ($m_{h^0}$) and $\sqrt{s}$ is plotted in Fig. \ref{fig:fig5} (right). It should be noted that, due to the small mass range for the light Higgs ($65<\mhl<120\gev$), the cross section changes very slowly with $\mhl$, and the difference is around -6\% at $\sqrt{s}=2\tev$ between the lower and the higher $m_{h^0}$ values. 
    Besides of these, the distribution of the cross section in Type-I and Type-II are the same in the numerical precision. When the exact alignment is chosen ($\sba=1$), the $\tb$ dependence of the cross section droped out over all in the calculation. The couplings $c_{e\bar{e}[h^0,H^0]}$ and $c_{\nu_e \bar{e}H^+}$ are a function of $tb$ via $\zeta_L$. Since the Yukawa scheme is relevant for the couplings beween Higgses and fermions, these couplings do not affect the calculation even varying at this range. Moreover, even if we lose the exact alignment and set $\sba=0.9$, the change in the cross section is around -2.5\% at $\sqrt{s}=1.5\tev$, and it falls at high COM energies. Finally, we get $\sigma_{\text{pol-1}}\approx 0.143\fb $ and $\sigma_{\text{pol-2}}\approx 0.175\fb $ for the polarized incoming beams.
    
\subsection{Favored Region by Recent Experimental Constraints}
 
\begin{figure}[htbp]
        \centering
        \includegraphics[width=\factor\textwidth]{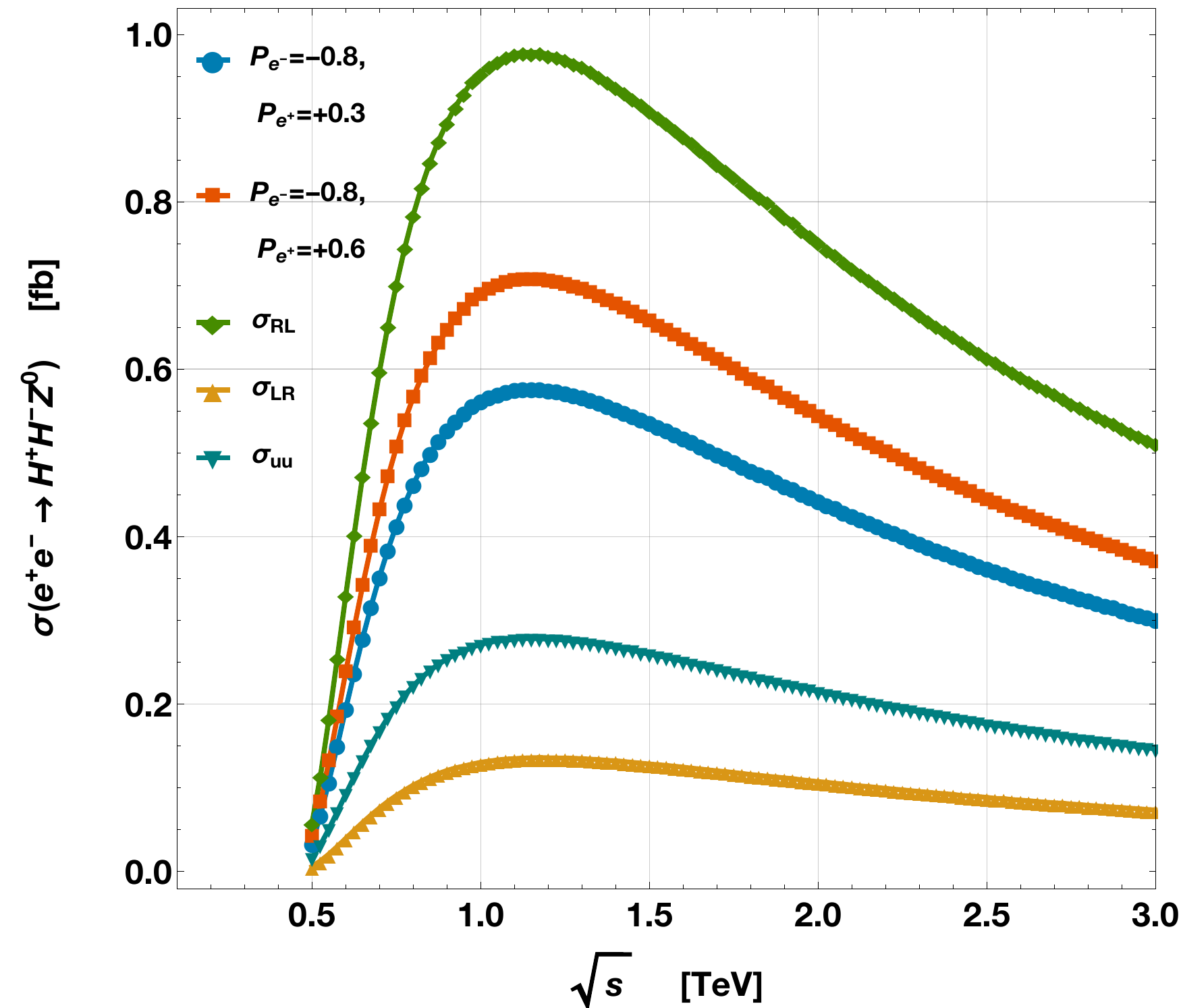}
        \includegraphics[width=\factor\textwidth]{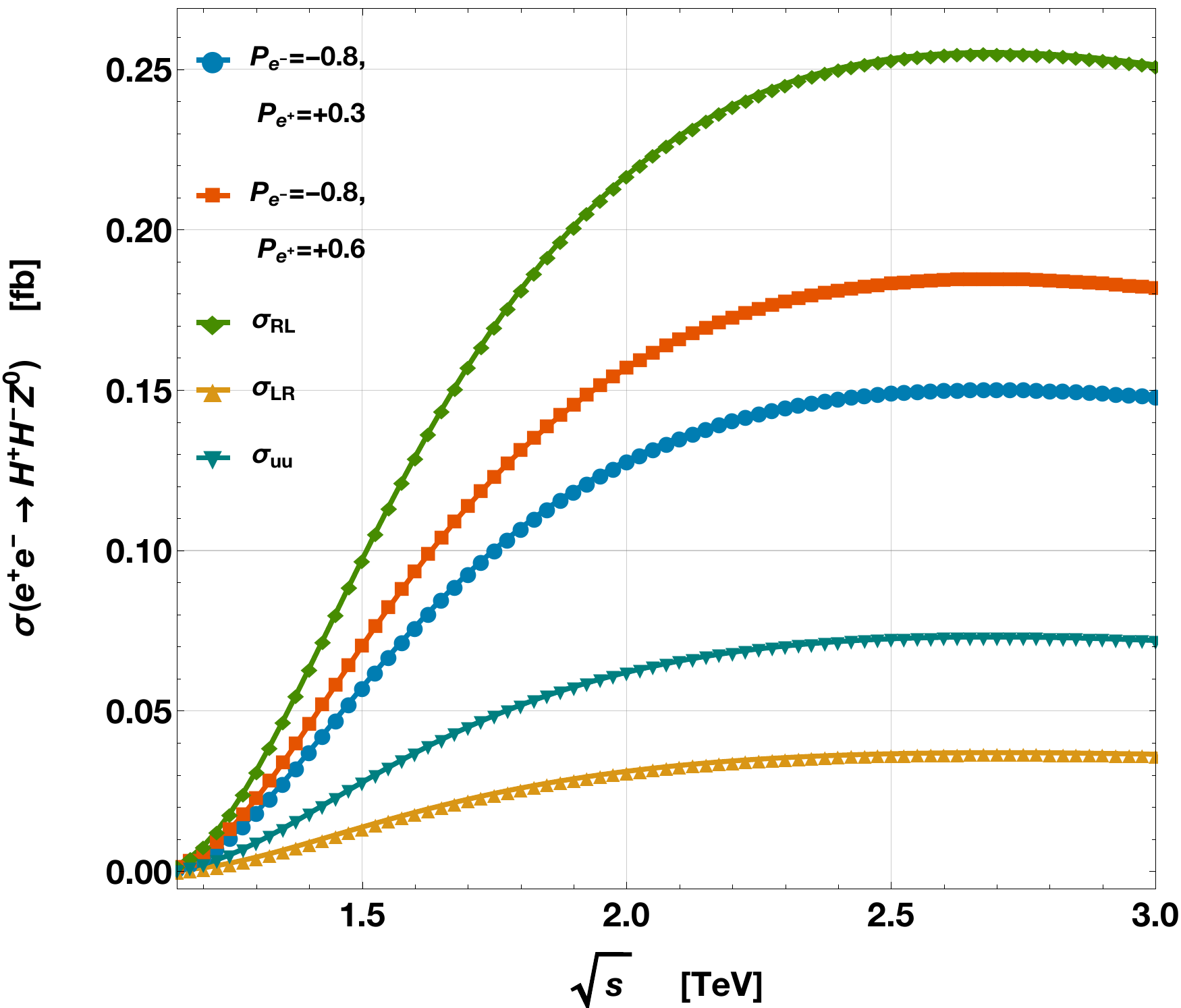}
        \caption{        \label{fig:fig6}
        The cross section distributions are plotted in favour of the current experiments in the flavour physics.
        (left): Cross section of the process as a function of $\sqrt{s}$ for various polarized beam configurations are plotted where Type-I Yukawa coupling scheme is set, $m_{H}=175 \gev$ and $\tb=10$,   
        (right): the same caption with the left figure but Type-II and $m_{H}=500 \gev$ are set.
        }
        \end{figure}

    The last set of analysis is presented considering the results obtained from the recent experiments particularly the flavor physics. In this scenario, there is no splitting between the charged Higgs mass, CP-odd and the heavier CP-even Higgs masses, ($m_{H}=m_{H^0}=m_{A^0}=m_{H^\pm}$). The light CP-even state is assumed to be the SM one, the exact alignment limit is used ($\sba=1$), and two parameters ($\tb$ and $\mhp$) given in Tab. \ref{tab:flavor_inspired} are regarded as a free parameters. The COM energy dependence of the cross section for unpolarized and polarized incoming beams are given in Fig. \ref{fig:fig6} up to $\sqrt{s}=3\tev$. On the left, the cross section for polarized beam configurations are plotted the same as the previous figures, Type-I Yukawa coupling scheme is set, all the extra Higgs masses are set $m_{H}=175\gev$, and $\tb=10$ is taken. According to \cite{Enomoto:2015wbn}, the restrictions on the charged Higgs mass in Type-I is much loose compared to the Type-II, and it is $150\gev < \mhp$. Due to the possibility of having a smaller charged Higgs mass the cross section gets much higher compared to the previous scenarios. The unpolarized cross section is calculated as $\sigma_{UU}\approx 0.278 \fb$, and the cross section ($\sigma_{RL}$) particularly for $P_{e^+e^-}=(+1,-1)$ goes up to $0.978 \fb$. The polarized cross section is obtained $\sigma_{\text{pol-1}}\approx 0.577 \fb$ and $\sigma_{\text{pol-2}}\approx 0.707 \fb$.

    The Type-II Yukawa coupling structure is set in Fig. \ref{fig:fig6} (right), and due to the restrictions from the flavor observables charged Higgs mass is constrained from below ($\mhp>500\gev$). The production rate is lowered compared to the Type-I, and the unpolarized cross section is $\sigma_{UU}=0.073 \fb$. However, the polarized cross section is obtained $\sigma_{\text{pol-1}}\approx 0.151 \fb$ and $\sigma_{\text{pol-2}}\approx 0.185 \fb$. In each case given in the Fig. \ref{fig:fig6}, polarization of the incoming beams enhances the cross section by a factor of 2.1 (2.5) for pol-1 (pol-2) compared to the unpolarized one, respectively.

\begin{figure}[htbp]
    \centering
    \includegraphics[width=\factor\textwidth]{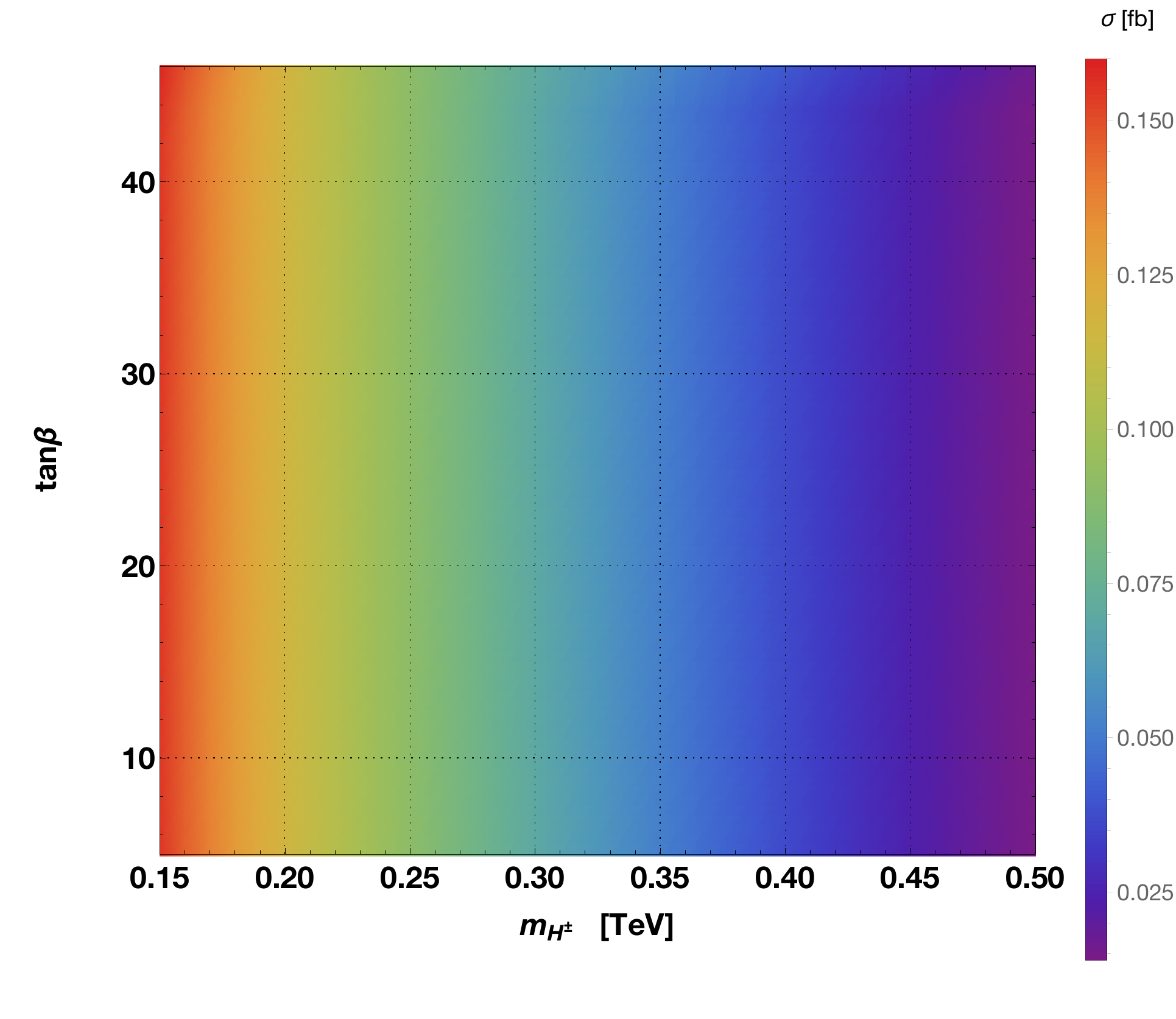}
    \includegraphics[width=\factor\textwidth]{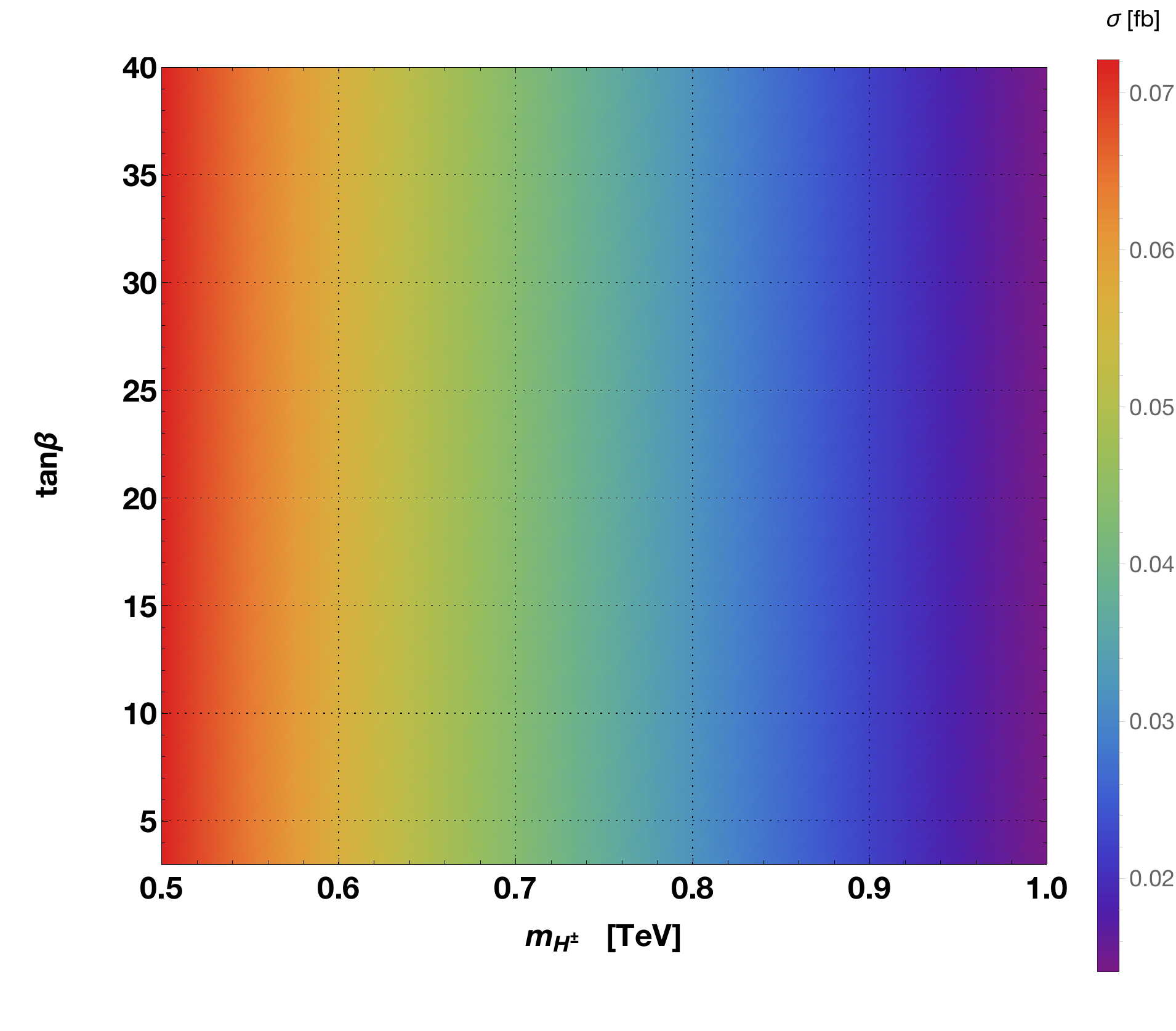}
    \caption{   \label{fig:fig7}
        The cross section distributions for the last scenario called favored region in the recent experimental constraints. The distributions are at  $\sqrt{s}=3\tev$, $\sba=1$ and all the extra Higgs masses are taken as $m_{H}=m_{H^0}=m_{A^0}=m_{H^\pm}$.
            (left): Type-I Yukawa couplings,  
            (right): Type-II Yukawa couplings are set.
    }
    \end{figure}

    The last result is given in Fig. \ref{fig:fig7} where the cross section at $\sqrt{s}=3.0\tev$ plotted as a function of $\tb$ and $\mhp$. On the left side, the Type-I Yukawa structure is set, and on the right, the type-II Yukawa structure is used. In Type-I, the cross section gets high for small charged Higgs masses, and it gets low as usual for high $\mhp$ values. However, as it can be seen clearly in both of the Fig \ref{fig:fig7}, $\tb$ does not affect the cross section. That means couplings given in Tab. \ref{tab:tab1} which have $\tb$ dependence indirectly through $s_\beta$, $c_\beta$ and directly in $\zeta_L$ do not influence the cross section in this special decoupling and exact alignment limit ($\sba=1$). Comparing the two plots at $\mhp=500\gev$ shows that the cross section in Type-I is higher by 25\%.
    

\section{Identifying the process at the detector}
\label{sec:6}

    In this section, the decay channels of the charged Higgs boson are discussed for all the three scenarios, and possible collider signatures for measuring the process are examined. To explore the process in a collider, primarily, we need to detect all the possible charged Higgs products. The charged Higgs boson decays through $\htb$ in all three scenarios commonly with varying branching ratios. The other channels are $\hwh$, $\hwH$ and $H^+\rightarrow \tau^+ \nu_\tau$. Besides of that, various differential distributions of the charged Higgs and $Z^0$ boson are calculated, and discussion is carried out whether they could be used for selection in the detector. Finally, the possible background channels and challenges are indicated for the detection of the process in the collider. According to \cite{Eidelman:2004wy}, $Z^0$ boson decays through three main channels with the following branching ratios; hadronic ($\approx 0.70$), leptonic ($\approx 0.10$), and invisible ($\approx 0.20$). In addition to that W-boson has hadronic ($\mathcal{B}(W\rightarrow \text{ hadronic})\approx 0.67$), and leptonic $\mathcal{B}(W\rightarrow l\nu_l) \approx 0.21$ decays (e and $\mu$).

\subsection{Decay channels of $H^\pm$ in the non-alignment scenario}

    In the non-alignment scenario, depending on the free parameters of the model, the charged Higgs boson is decayed mainly in the following decay channels, $\htb$ and $\hwH$ in Type-I, and another channel is joined to them in Type-I which is $\hwh$. In Fig. \ref{fig:fig8} and \ref{fig:fig9}, the branching ratios in the non-alignment scenario are drawn as a function of $m_H$ and $\tb$ for each Type-I and -II, respectively. It can be seen that the decay of $\hwh$ is going down while $\hwH$ becomes dominant. A seesaw between the decays of $\hwh$ and $\hwH$ appears.  $\mathcal{B}(\hwH)$ is greater than 90\% in region $m_H\lesssim350\gev$ and $\tb\gtrsim6$. Then, it gradually decreases at high $m_H$ values, and the decay channel $\hwh$ gains weight. 

        \begin{widetext}
        \onecolumngrid
    \begin{figure}[htbp]
    \centering
    \includegraphics[width=\factortwo\textwidth]{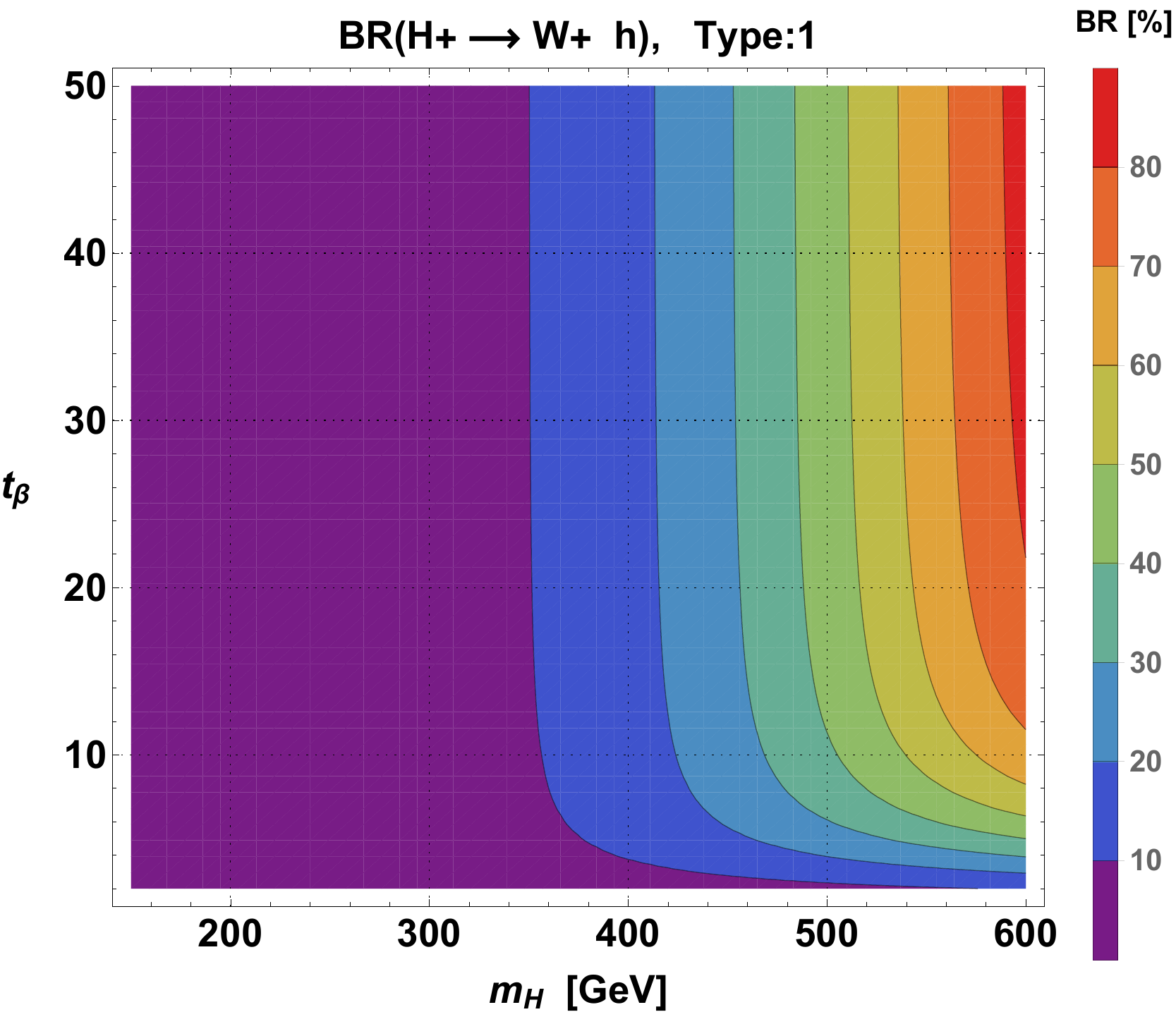}
    \includegraphics[width=\factortwo\textwidth]{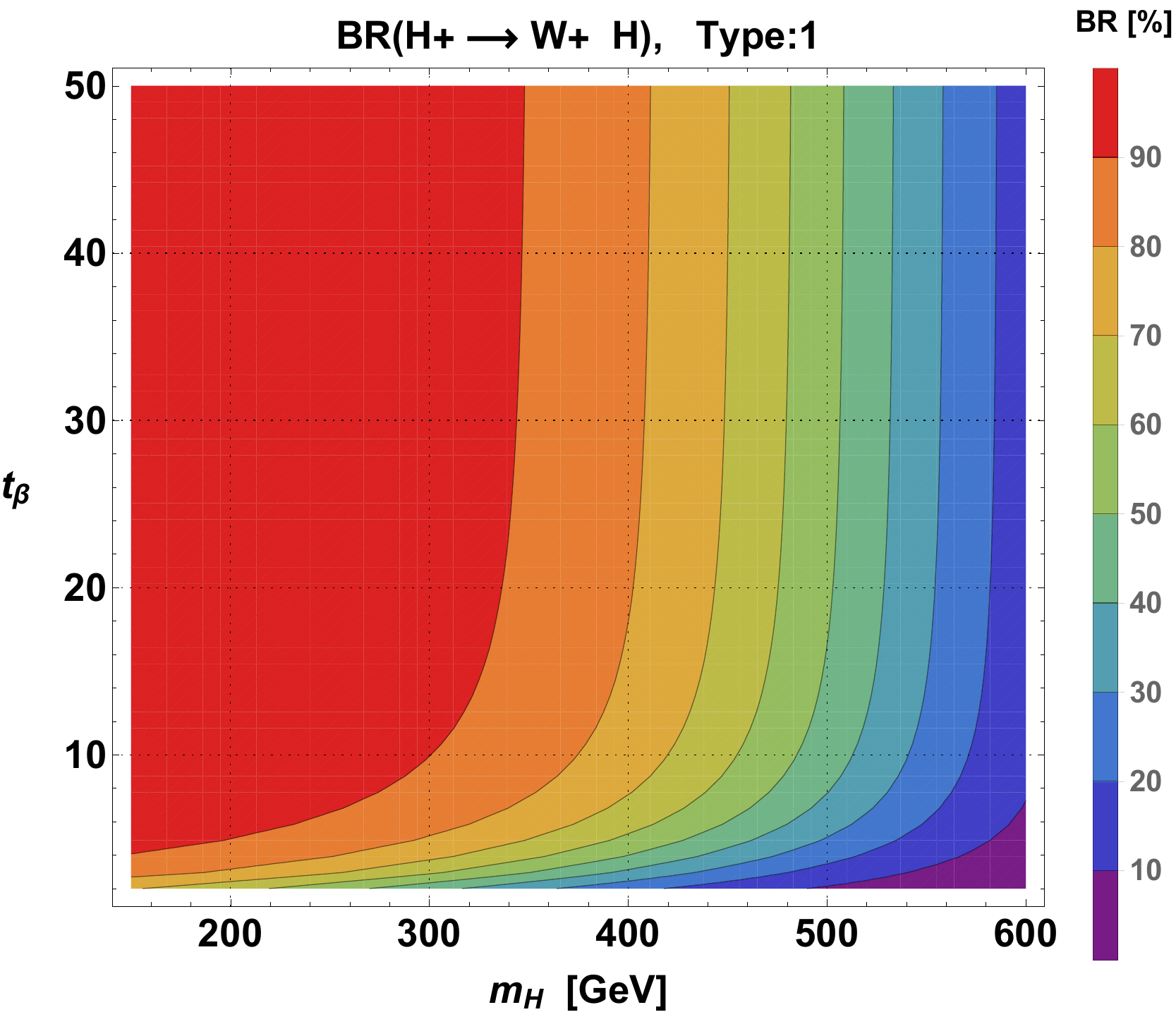}
    
    \includegraphics[width=\factortwo\textwidth]{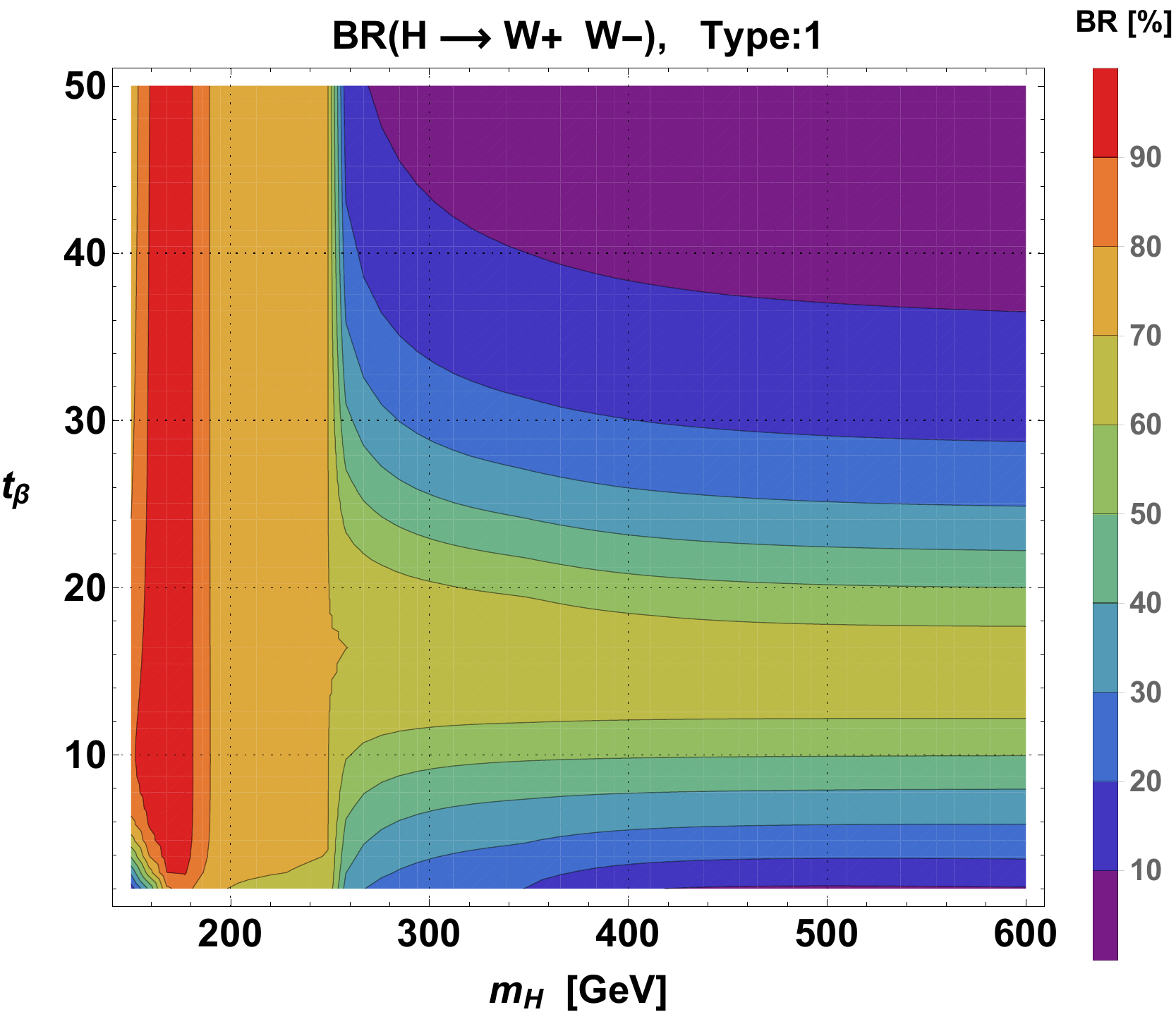}
    \includegraphics[width=\factortwo\textwidth]{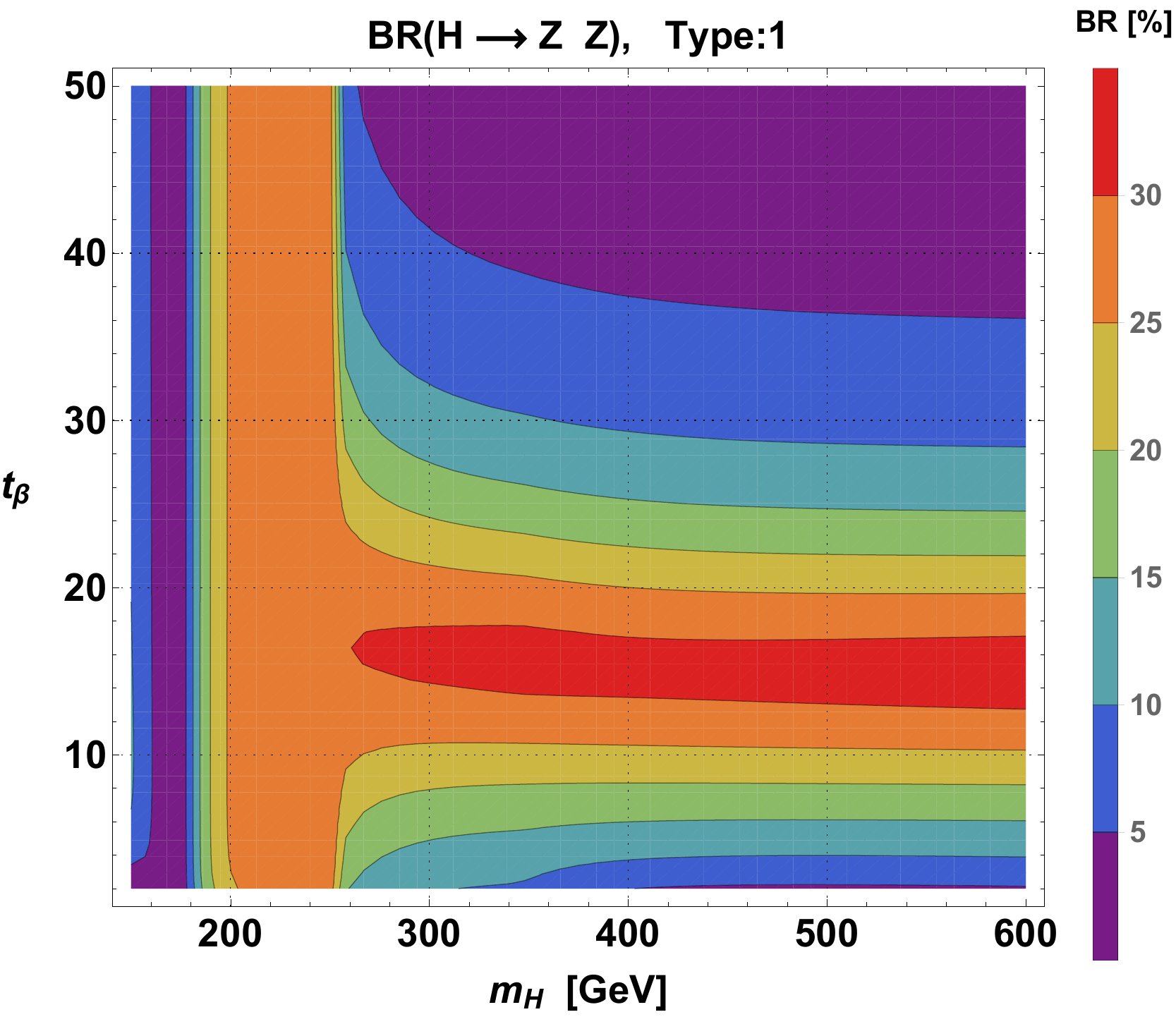}
    \includegraphics[width=\factortwo\textwidth]{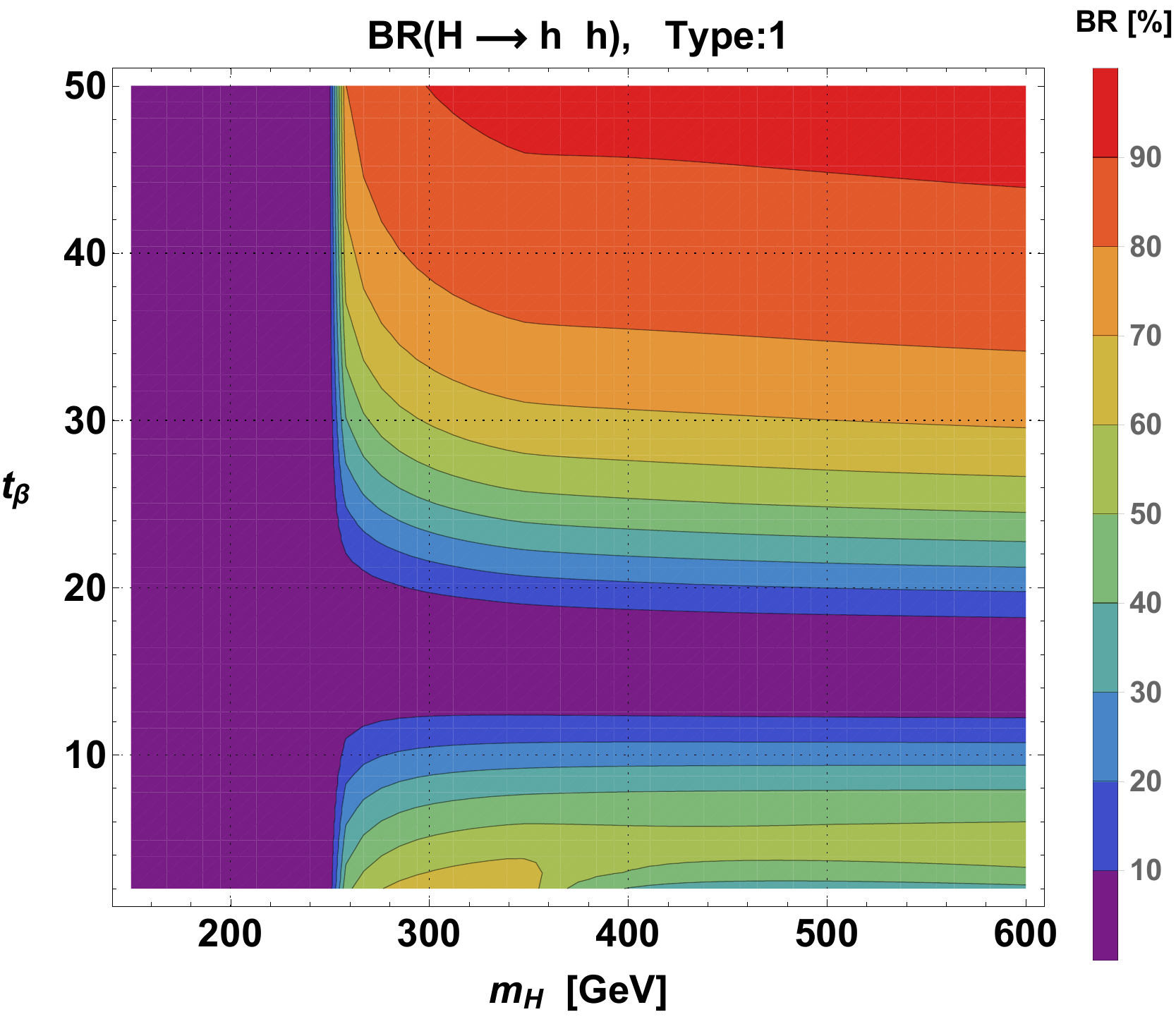}
    \caption{ \label{fig:fig8}
        Distributions are plotted for the branching ratios of the charged Higgs boson and CP-even $H^0$ boson in the non-alignment scenario. 
        Decay channels are indicated in each figure.
        All the extra Higgs masses are taken as $m_{H}=m_{A^0}=m_{H^\pm}$, and $m_{H^0}$ is calculated with the help of Eq. \ref{eq:eq10}.
        Type-I Yukawa couplings are assumed, and $\cba=0.1$ is set.
        (first row): Decays of $H^\pm$ are plotted.
        (second row): Decays of CP-even $H^0$ boson are plotted.
    }
    \end{figure}
    \end{widetext}
    In the non-alignment scenario with Type-I, the process becomes as $e^+e^-\rightarrow H^+H^-Z\rightarrow W^+h^0 W^-h^0 Z$, and the hadronic decays of W and $Z^0$ bosons could be an ideal option for reconstructing the process at $m_H\gtrsim 500\gev$ given in Fig. \ref{fig:fig8}. Since, $h^0$ decays mostly through $\mathcal{B}(h^0\rightarrow b\bar{b})\approx62\%$ and $\mathcal{B}(h^0\rightarrow W^+W^-)\approx 20\%$, and considering the $Z\rightarrow q\bar{q}$, the final state of the process in this region will be 4 jets + 4 b-tagged jets + 2 jets (coming from $Z^0$ boson). The heavier CP-even Higgs boson with various decay channels are given in Fig. \ref{fig:fig8} (second row).  The charged Higgs decays through the heavier CP-even Higgs boson and W-boson for $m_H\lesssim500\gev$, and then $H^0$ most likely decays to vector boson pairs ($WW/ZZ$) or $h^0h^0$ pairs. Considering the hadronic decays of the vector bosons and $h^0$, for the low $m_H$ region, there will be in total 12 jets at the final state and also the decay products of the $Z^0$ boson (2 jets or 2 leptons). Apparently, this scenario produces many jets at the final state, and it is a challenge to reconstruct the W-boson, so the charged Higgses. The performance of the jet finding algorithms in such a jetty environment is vital to study the production and the couplings of the charged Higgs. Non-alignment with Type-I has a potential to produce the highest number of jets in the final state among all other cases presented in this study. 
    
        \begin{widetext}
        \onecolumngrid
        \begin{figure}[htbp]
        \centering
        \includegraphics[width=0.40\textwidth]{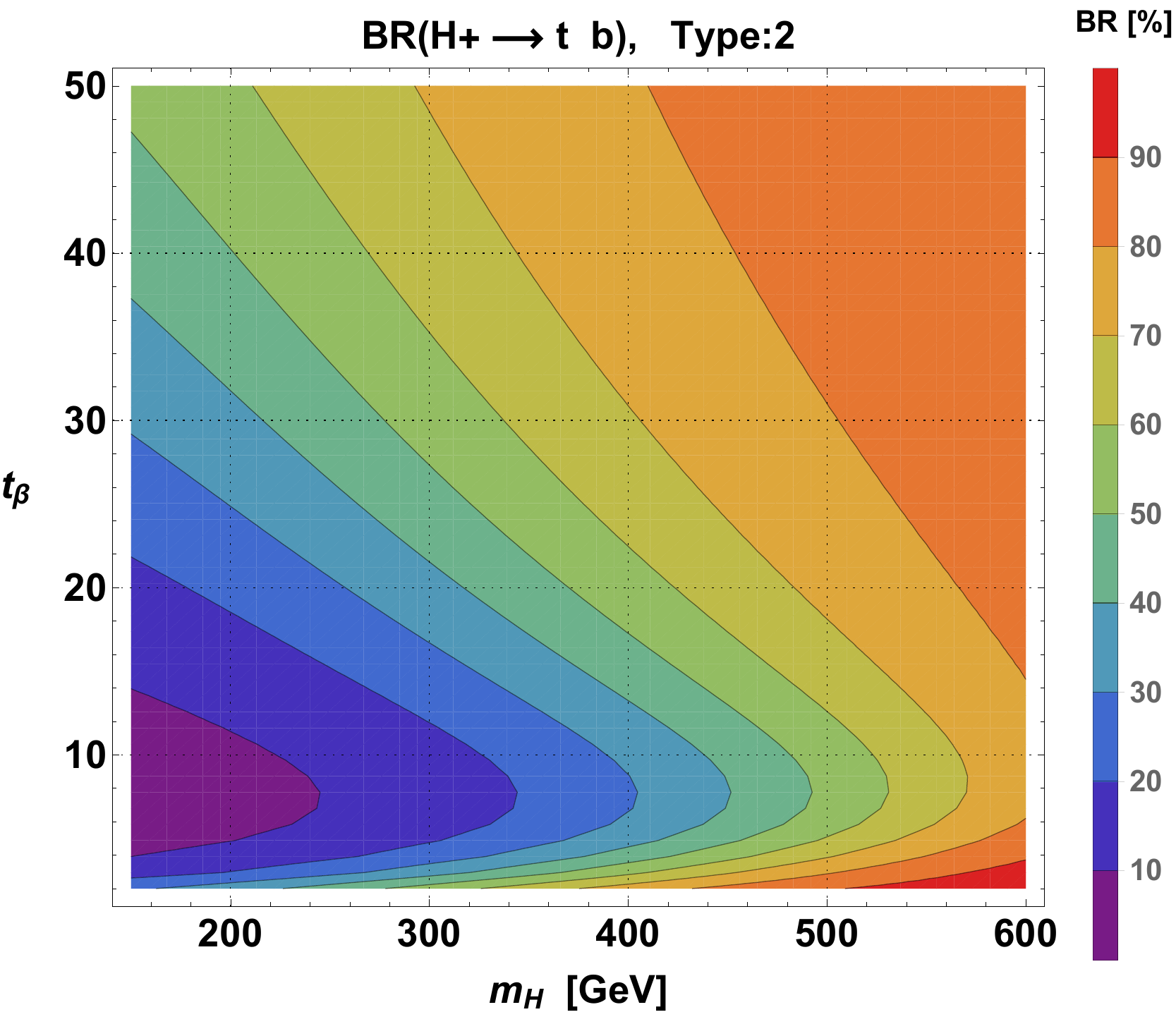}
        \includegraphics[width=0.40\textwidth]{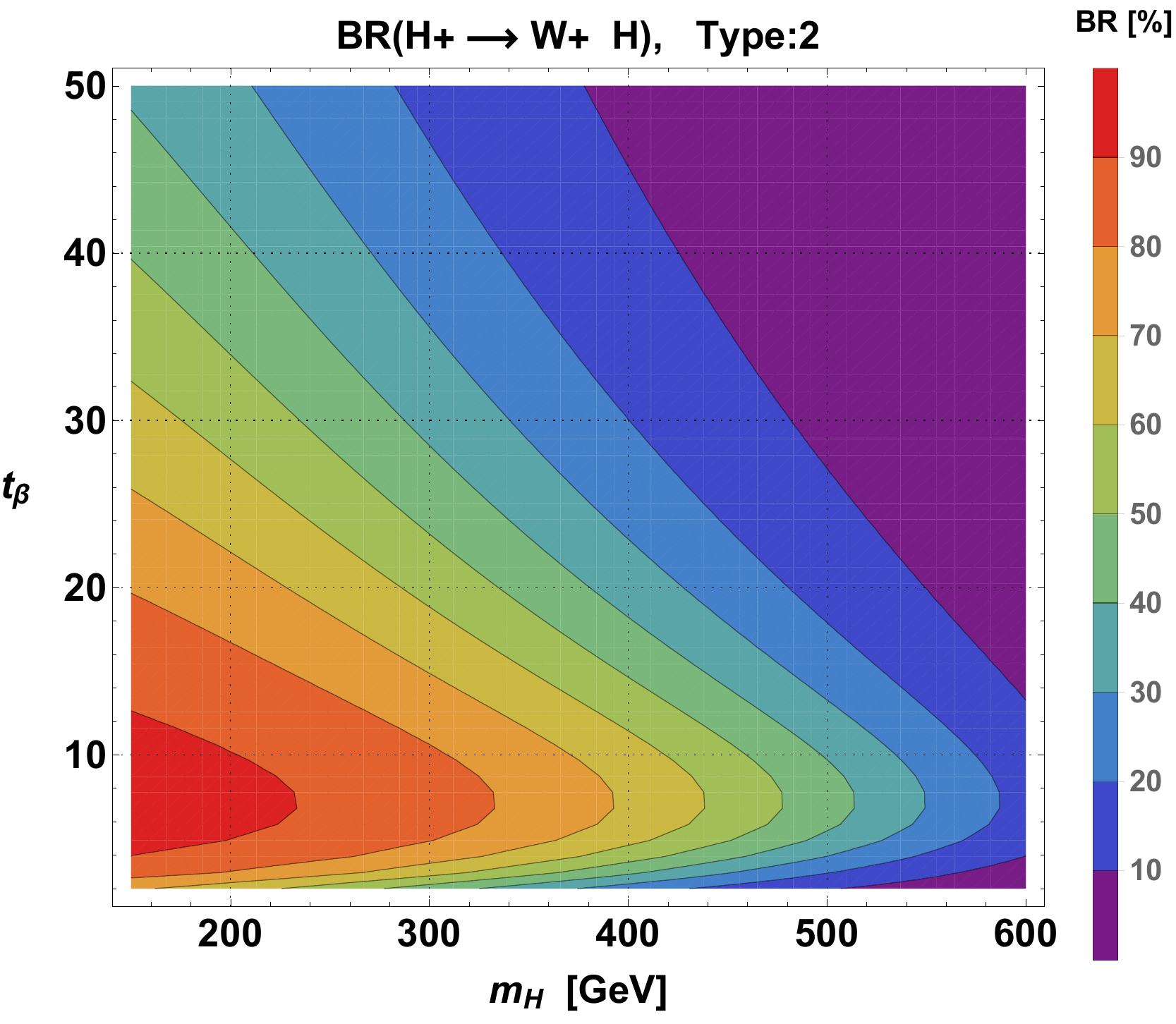}
    \caption{\label{fig:fig9}
    Distributions for the branching ratios of the charged Higgs boson in the non-alignment scenario. 
    Decay channels are indicated in each figure.
        The Higgs masses are taken as $m_{H}=m_{A^0}=m_{H^\pm}$, and $m_{H^0}$ is calculated with the help of Eq. \ref{eq:eq10}.
    The calculation is carried for $\cba=0.01\left( \frac{150\gev}{m_{H^0}}\right)^2$, and    Type-II Yukawa couplings are assumed.
    }
    \end{figure}
    \end{widetext}
    The situation is less complicated in Type-II, the charged Higgs decays through $\htb$ at high $m_H$ and $\tb$ values, and the decay channel $\hwH$ gains weight at low $m_H$ and $\tb$ region given in Fig. \ref{fig:fig9}. Besides, $\mathcal{B}(H^0\rightarrow b\bar{b})\gtrsim 90\%$ at the most of the parameter space. An ideal case for this scenario would be letting $H^0$ decays through $b\bar{b}$ quarks and hadronic decay of the $W^\pm/Z^0$ bosons. Then, there will be 4 b-tagged jets + 6 jets in the final state. 
    Additionally, it is possible to trigger the events with $Z\rightarrow l\bar{l}$, and there will be two leptons with opposite sign in the final state. Unfortunately, the branching ratio of the leptonic decays of the $Z^0$ boson is small compared to the hadronic decays. 
    The decay channel $\htb$ is significant at high $m_H$ values. Then, the task is to look for top quarks in the final state.
    Consequently, the subsequent decays of $t\rightarrow W b$, $W^\pm\rightarrow q\bar{q}(l\nu_l)$ will form the signature of the charged Higgs boson at a detector.
    If we let $Z\rightarrow q\bar{q}(l\bar{l})$, the process $\theprocess\rightarrow tb tb Z$ could be reconstructed with 10 jets + 2 b-tagged jets or 8 jets + 2 b-tagged jets+ 2 leptons, respectively.
    It is seen that tagging the b-quark and reconstructing the charged Higgs mass with the possible decay products of the W-boson is vital for the process.
    Another choice, which is common in the next scenarios as well, is to let W decay leptonically and instead of two jets in the final states there will be a lepton + missing $E_T$. However, Monte Carlo simulation study would be the best to determine the efficiency of the leptonic decay channel. 
    
\subsection{Decay channels  of $H^\pm$ in the low-$m_H$ scenario}

\begin{figure}[htbp]
        \centering
        \includegraphics[width=0.40\textwidth]{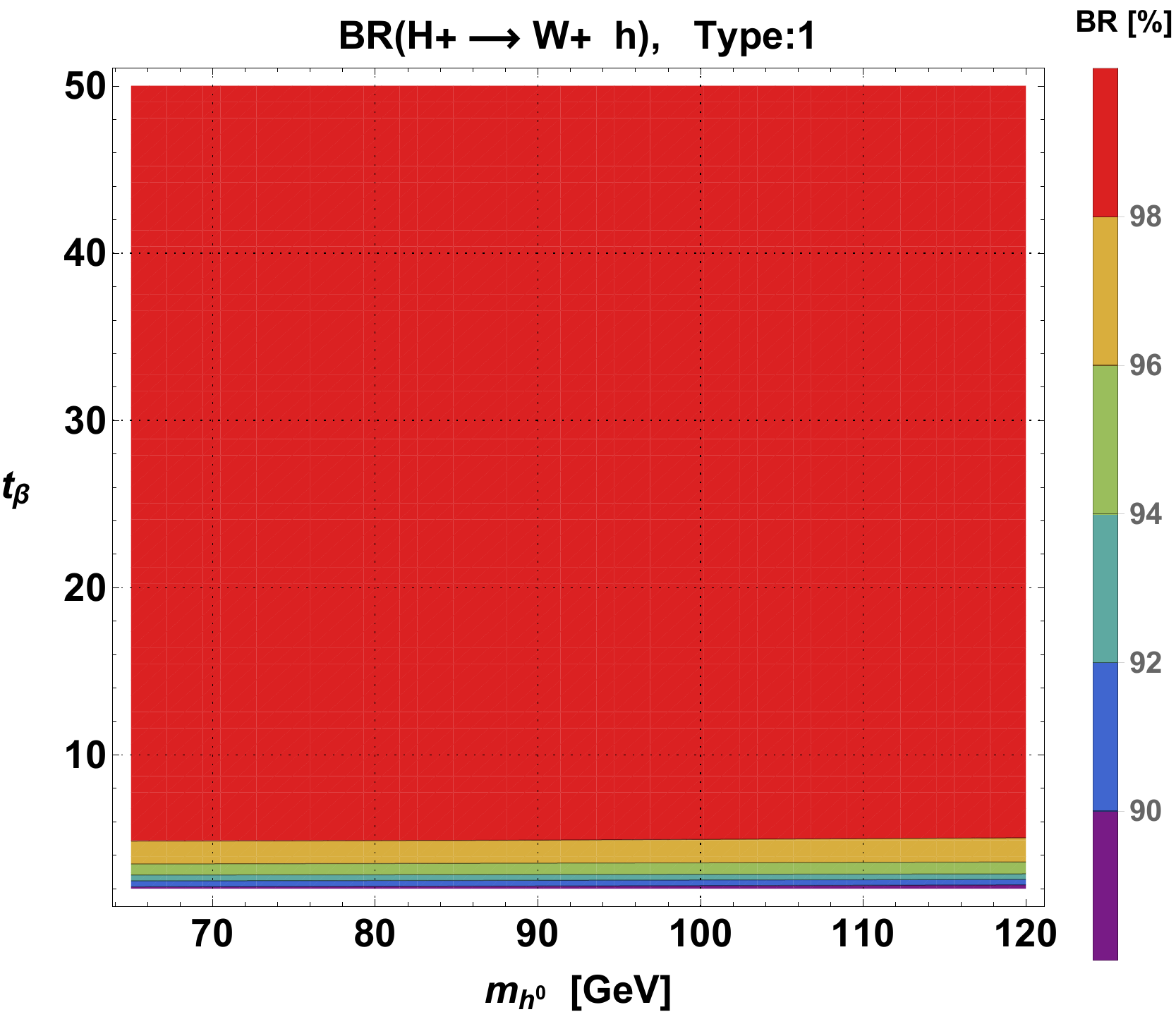}
        \includegraphics[width=0.40\textwidth]{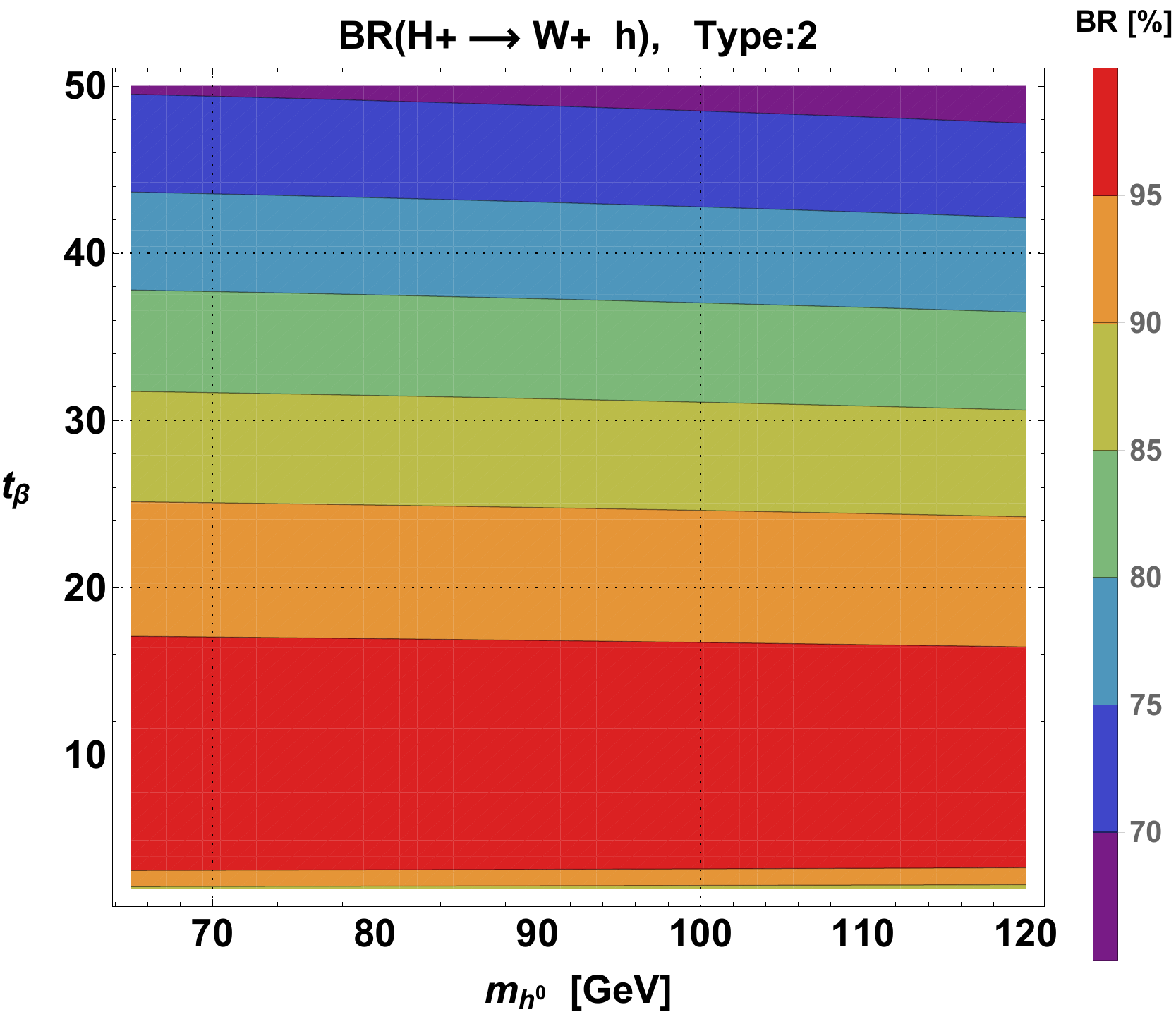}
        \caption{\label{fig:fig10}
            Distributions for the branching ratios of the charged Higgs boson in the low-$m_H$ scenario. 
            Distributions are given for $\cba=1$, $m_{H^0}=125\gev$, $\maa=\mhp$ are calculated with the help of Eq. \ref{eq:eq10}.
            (left): Type-I Yukawa couplings are assumed,  
            (right): Type-II Yukawa couplings are assumed.
        }
        \end{figure}

    In the low-$m_H$ scenario, the mass gap between the CP-even Higgs bosons is small, and the charged Higgs boson decays via two channels $\hwh$ and $\htb$ in both Type-I and -II. The branching ratios for each type are given In Fig. \ref{fig:fig10}, and the sum of $\mathcal{B}(\hwh)$ and $\mathcal{B}(\htb)$ adds up to unity in this scenario. It is $\mathcal{B}(h^0\rightarrow b\bar{b})\gtrsim 80\% (90\%)$ in Type-I (Type-II), respectively. If $h^0\rightarrow b\bar{b}$ is considered, the process could be tagged with 4 b-tagged jets + 4 jets + $Z^0$ boson in both of the Yukawa coupling schemes (Type-I/-II).

\subsection{Decay channels of $H^\pm$ in the favored region}

\begin{figure}[htbp]
        \centering
        \includegraphics[width=0.40\textwidth]{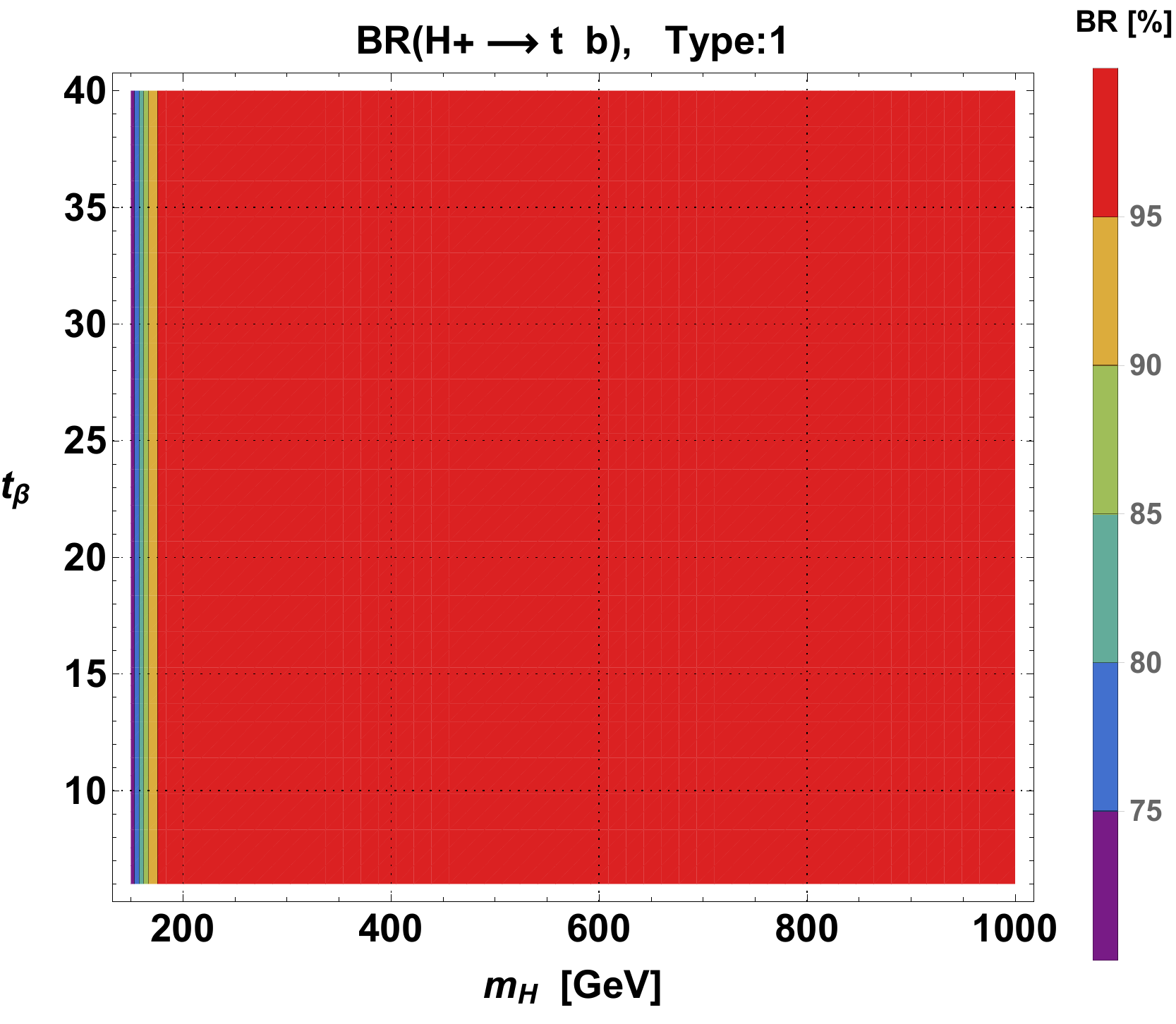}
        \includegraphics[width=0.40\textwidth]{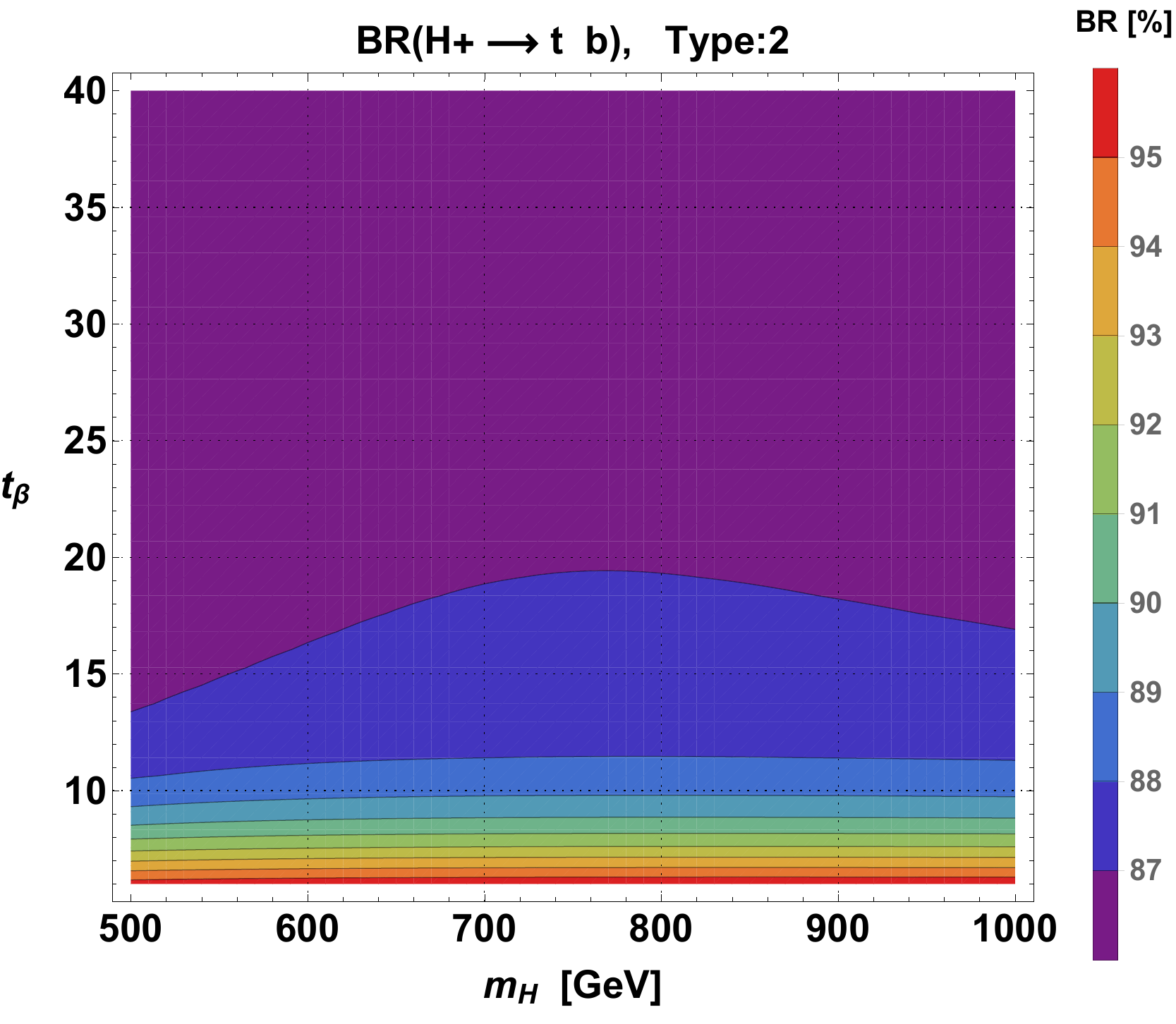}
        \caption{\label{fig:fig11}
        Distributions for the branching ratios of the charged Higgs boson in the favored region by the recent experimental constraints.
        Distributions are for $\sba=1$, and all the extra Higgs masses are taken as $m_{H}=m_{H^0}=m_{A^0}=m_{H^\pm}$.
        (left): Type-I Yukawa couplings are set,  
        (right): Type-II Yukawa couplings are set.
        }
        \end{figure}

    In this scenario, the charged Higgs boson decays via two channels $\htb$ and $H^+ \rightarrow \tau \nu_\tau$ for both Type-I and -II.
    In Fig. \ref{fig:fig11}, it can be seen that the $\mathcal{B}(\htb)$ is the dominant decay channel for each Yukawa structure.
    Then, the same final state will be obtained with the non-alignment scenario Type-II.
    The subsequent decays of $t\rightarrow W b$ and $W^\pm\rightarrow q\bar{q}(l\nu_l)$ forms the final state of the charged Higgs boson.
    Then, considering the $Z\rightarrow q\bar{q}$, the final state of the process is 4 jets + 4 b-tagged jets + 2 jets.
    
 \subsection{Differential distributions} 

\begin{figure}[htbp]
        \centering
        \includegraphics[width=\factortwo\textwidth]{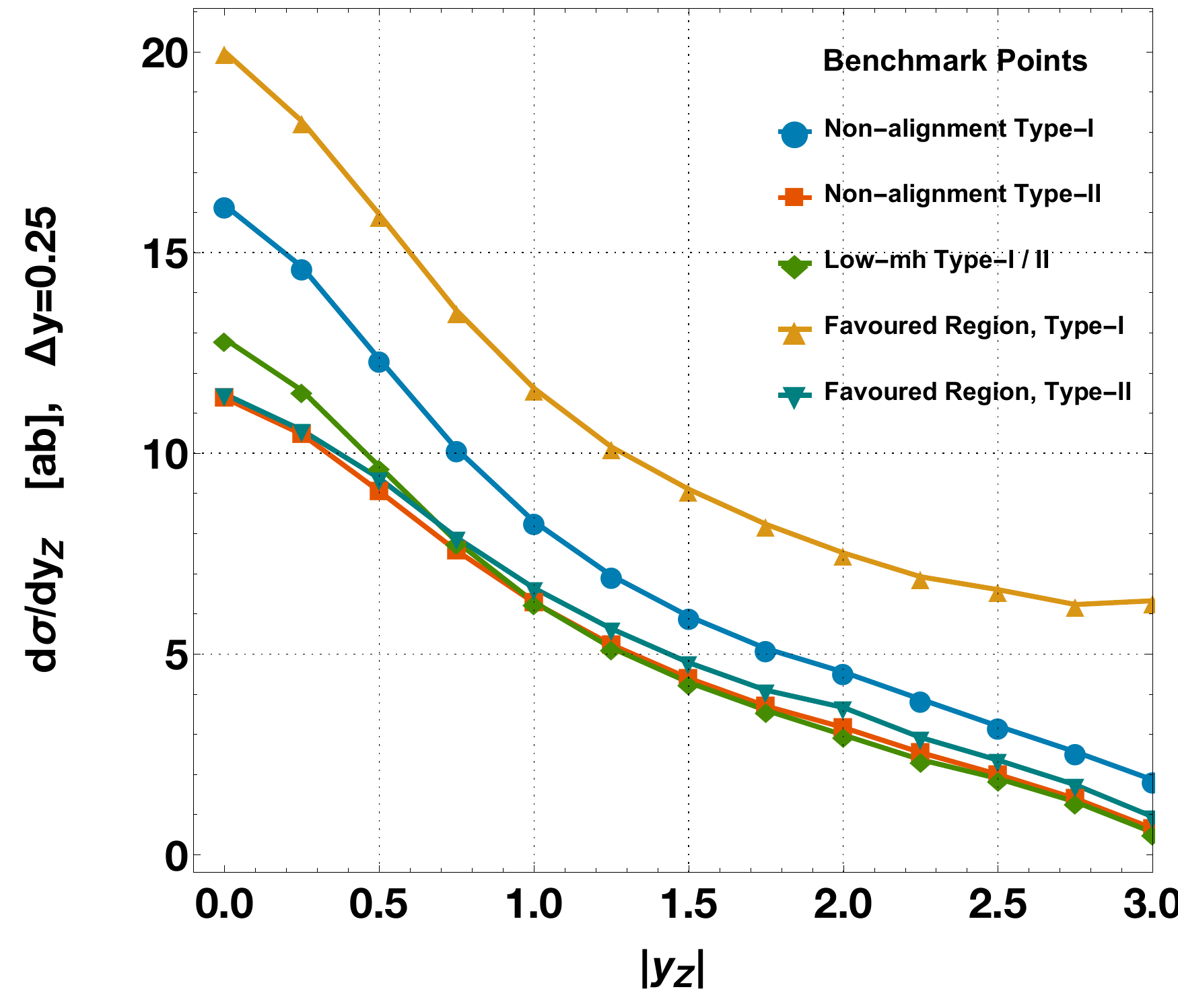}
        \includegraphics[width=\factortwo\textwidth]{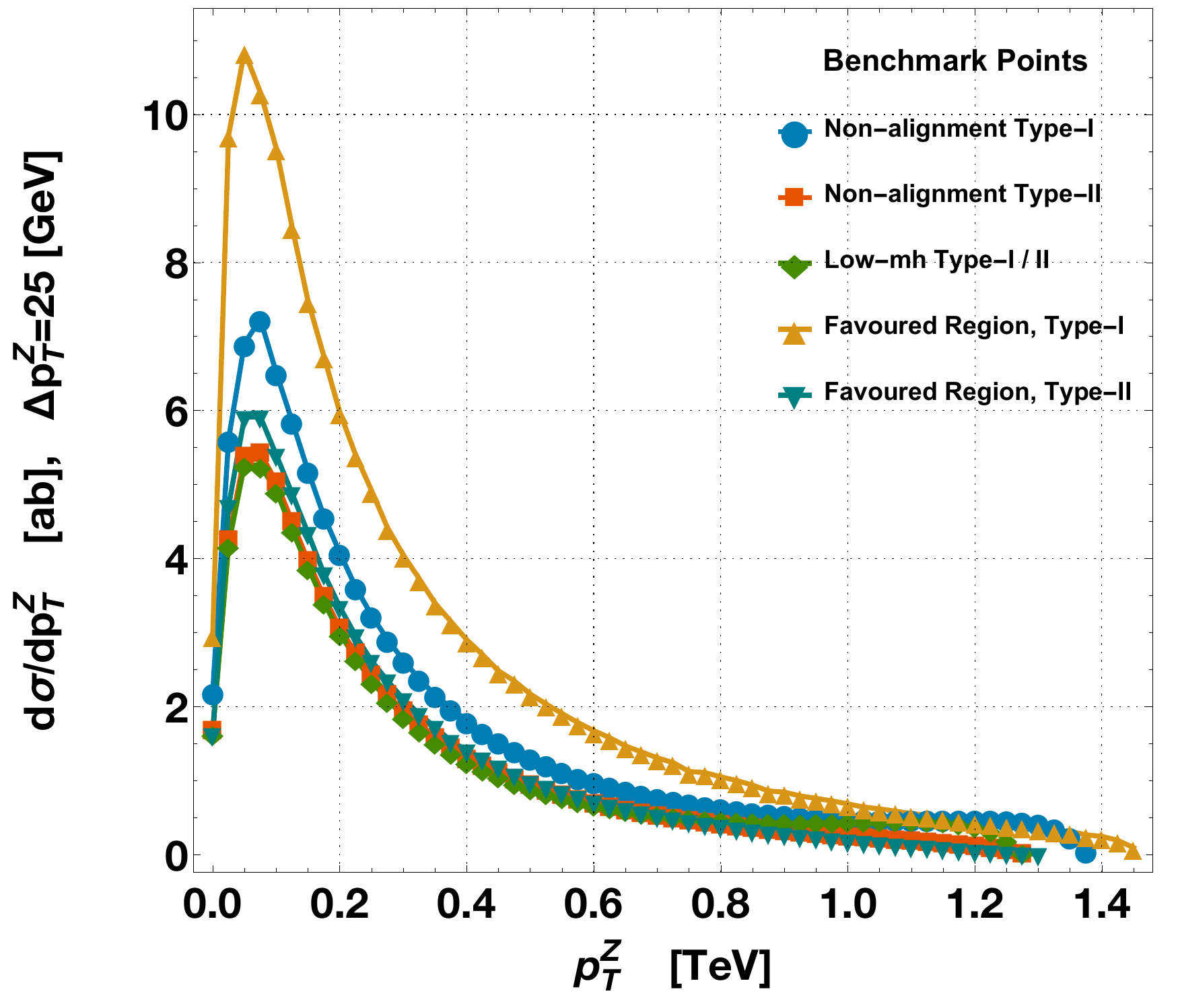}
        \includegraphics[width=\factortwo\textwidth]{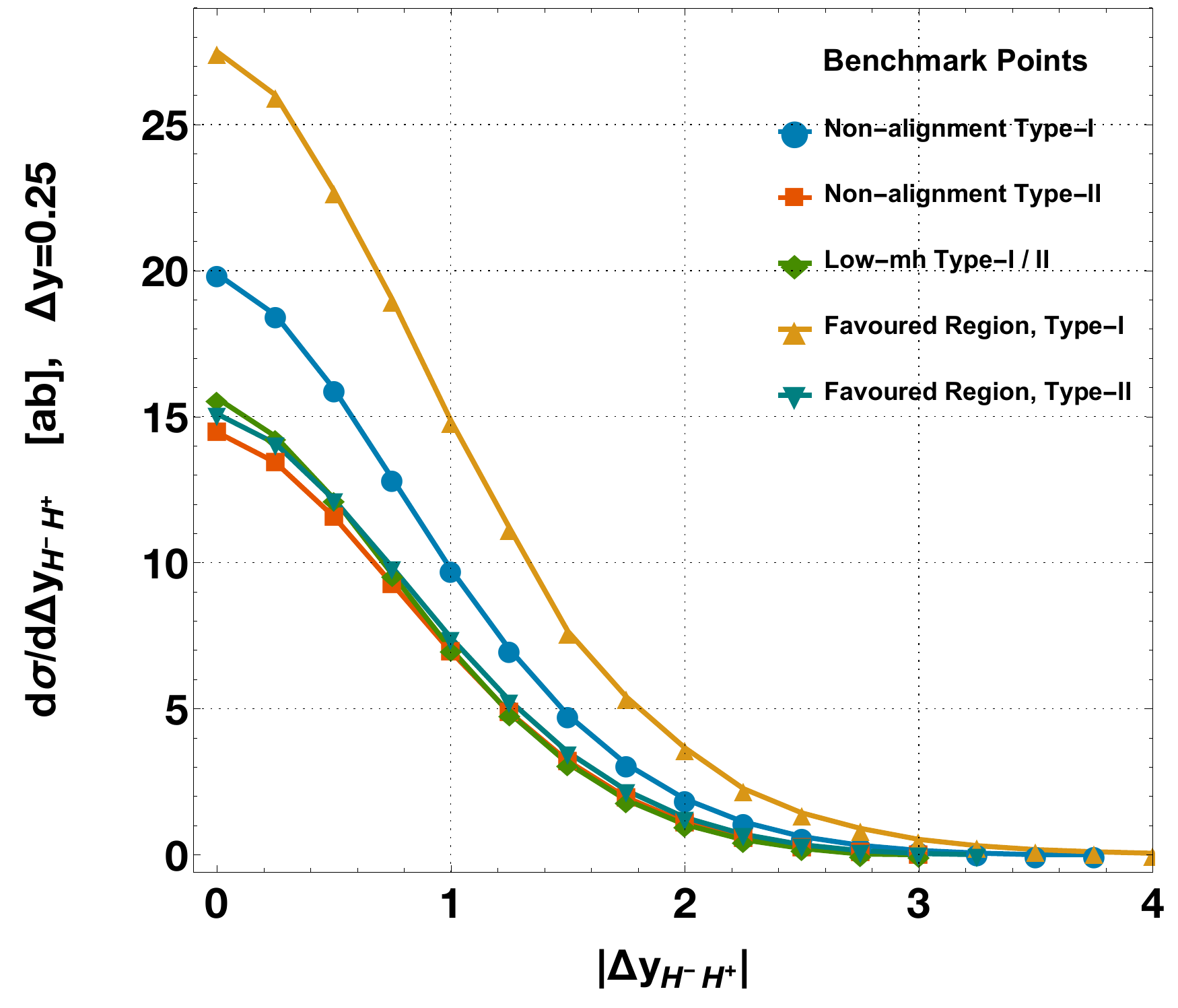}
        \caption{ \label{fig:fig12}
            The differential cross section as a function of the kinematical properties of Z and $H^\pm$ bosons where $\sqrt{s}=3\tev$, and the rest of the parameters are indicated in the text for each scenario.
            (left):   $d\sigma/d|y_Z|$ distribution,  
            (center): $d\sigma/dp_T^Z$ distribution,
            (right):  $d\sigma/d|\Delta y_{H^-H^+}|$ distribution.
        }
        \end{figure}

    In this section, the differential cross sections are calculated as a function of the kinematical variables for each scenario, and comparison is performed. The computation is carried at $\sqrt{s}=3\tev$, $\sba$ or $\cba$ is taken from the corresponding table given for the each scenario. Higgs masses are set to $\mhh=425\gev$ ($\tb=45$) in the non-alignment scenario, $\mhl=80\gev$ in the low-$m_H$ scenario, finally, $m_{H}=m_{H^0}=m_{A^0}=m_{H^\pm}=175 \;(500)\gev$ for Type-I (Type-II) in the favored region, respectively. The $\tb=10$ is set for all the scenarios.
    In Fig. \ref{fig:fig12} (left), the differential cross section as a function of the rapidity of the $Z^0$ boson is presented. It can be seen that the $Z^0$ boson is produced more likely in the central region compared to the high rapidities for all the scenarios. If a rapidity cut of $|y_Z|\leq2$ on the $Z^0$ boson is applied, then depending on the scenario $\approx 75-88\%$ of the events could be captured. That shows it could be a useful kinematical cut to eliminate more events in the background. 
    In Fig. \ref{fig:fig12} (center), distribution for the differential cross section as a function of the transverse momentum of the $Z^0$ boson ($p_T^Z$) is plotted for each scenario. According to the $p_T^Z$ distribution, the $Z^0$ boson is produced more likely with small transverse momentum. That will most certainly affect the kinematical properties of the decay products of the $Z^0$ boson. Applying a cut of $p_T^Z<400\gev$ reaps more than $\approx70\%$ of the events.
    At last, the differential rate as a function of the rapidity difference of the charged Higgses is plotted in Fig. \ref{fig:fig12} (right), and it shows that the charged Higgses are produced very likely with a small rapidity difference, and $\approx95\%$ of the events fall in $|\Delta y_{H^- H^+}|\leq2$. Moreover, the normalized rate is the same in all the scenarios. 
    
 \subsection{SM Backgrounds}

    The production of $\products$ in $e^+e^-$ collider has a small cross section as it is presented in the previous section. Regarding the weakness of the signal, one may wonder if such a process could be extracted from the SM background. The decay chains of the charged Higgs in each scenario showed that the b-quark identification is vital in the reconstruction of the signal. Besides, the number of jets and the b-tagged jets at the final state in real-world are not fixed due to the parton-branching of the quarks and the efficiency of the b-tagging as well as the mis-identification of them. By considering these points, there are many background channels which will shadow the process. Therefore, a large background is expected mainly from $e^+e^-\rightarrow q \bar{q}$, and where later quarks will hadronize to jets. Moreover,  
        $e^+e^-\rightarrow  q \bar{q} Z $,
        $e^+e^-\rightarrow  t \bar{t}   $, and 
        $e^+e^-\rightarrow  t \bar{t} Z $ will contribute to the main background. On the other hand, the following processes could also contribute to the backround. These are $e^+e^-\rightarrow t\bar{t}b\bar{b}b\bar{b}$ along with $Z^0$ boson or multiple numbers of light jets (u, d, c, or s type quark), 
        $e^+e^-\rightarrow t\bar{t}ZH$, 
        $e^+e^-\rightarrow t\bar{t}ZZ$, 
        $e^+e^-\rightarrow t\bar{t}HH$, 
        $e^+e^-\rightarrow t\bar{t}b\bar{b}Z$, and 
        $e^+e^-\rightarrow W^+HW^-HZ$.        

    Hadronic decays of the $W^\pm/Z^0$-boson along with the jets could mimic the final state of the process. It looks like the top quark is essential for reconstructing the charged Higgs as well as in the elimination of the various background channels. Besides, it is also possible to allow the leptonic decays of $W^\pm/Z^0$ bosons. However, since the branching ratios are small, the multiplication of $\mathcal{B}(W\rightarrow l\nu_l)^2\cdot\mathcal{B}(Z\rightarrow l\bar{l})\approx 4.4\cdot 10^{-3}$ reduces the cross section by $\sim1/225$. Eventually, exploring the hadronic decays of $W^\pm/Z^0$ bosons gives more events in the detector. Therefore, jet finding algorithms will determine whether the process could be measured due to the challenges in the high jet multiplicity. The full reconstruction of W, t, $H^\pm$\, and better efficiency of b-tagging certainly boost the discrimination power of extracting the signal from the background. The observability of the process $\theprocess$ would require Monte Carlo simulation of the signal and all possible background processes, which is beyond the scope of this paper.

\section{Conclusions}
\label{sec:7}

    In this paper, we have calculated the process $\theprocess$ at the tree level in 2HDM. The process is analyzed for three scenarios motivated by the current experimental constraints. They are the non-alignment scenario, the low-$m_H$ scenario, and the scenario inspired from the results in the flavor physics. In the non-alignment scenario, the unpolarized cross section has the value of $0.071\;(0.106)\fb$ depending on the Yukawa coupling scheme Type-II (Type-I), respectively. In the low-$m_H$ scenario, the unpolarized cross section gets a value of $0.071\fb$ and contrary to the other scenarios Yukawa coupling scheme does not have an impact. In the last scenario, which is motivated by the observables measured in the flavor physics, the unpolarized cross section gets a value of $0.279\;(0.073)\fb$ depending on the Yukawa coupling schemes Type-I (Type-II), respectively. Comparing each scenario shows that the cross section is enhanced to a factor of 2-2.5 depending on the scenario for left-handed polarized electron beam ($P_{e^-}=-0.80$) and right-handed polarized positron beams ($P_{e^+}=+0.60$). The option of upgrading the incoming electron and the positron beam to be polarized has the power to enhance the potential of the machine. If it is assumed that CLIC could produce a total of integrated luminosity of $\mathcal{L}\sim 3 \text{ ab}^{-1}$ at $\sqrt{s}=3\tev$ \cite{CLIC:2016zwp}, then there could be more than $\sim2.7\cdot10^3$ events assuming polarization.

    The decay channels of the charged Higgs boson in each scenario are also investigated with the help of \texttt{2HDMC}.  The analysis shows that some channels come forward such as $\htb$, $\hwh$, and $\hwH$. The subsequent hadronic decays of the top quark, $W^\pm/Z^0$ bosons, and $h^0/H^0$ bosons have higher branching ratio compared to the leptonic ones. In that case, the final state of the process contains 6 jets + 4 b-tagged jets in the non-alignment scenario with Type-I, and 6 jets + 8 b-tagged jets in Type-II. Moreover, the other two scenarios have the same final state though following different decays, and there are 6 jets + 4 b-tagged jets in the end. Unfortunately, since the number of jets in the final state is high, it will be hard to reconstruct the process, and that is the most significant disadvantage. High efficiency in b-tagging and reconstructing the $W^\pm/Z^0$ boson, then reconstructing the top quark is vital for the charged Higgs detection. Further, charged Higgs pair has higher production rate compared to the extra $Z^0$ boson in the final state. Considering the background, a detailed Monte Carlo study is required to determine the significance and the acceptance of the signal in a detector. Some differential distributions for the charged Higgs and $Z^0$ boson are presented, but the best selection cuts with a higher elimination of the background signals require a full detector simulation and maybe a help of the artificial neural networks.
    
    LHC experiment confirmed the existence of a neutral Higgs boson \cite{Aad:2012tfa, Chatrchyan:2012xdj, Khachatryan:2014jba, Aad:2015gba}, hereafter, the discovery of another scalar and even a charged one at the future colliders would be clear evidence of the existence of new physics beyond the SM. In recent years, LHC experiment provided many results but no hint of the new physics yet. Since the precision measurements on Higgs and extended Higgs sector is a primary motivation for the LCC, this study shows the ability of the measurements for the process and particularly the charged Higgs sector of the 2HDM. Studying the production of $\products$ also revealed that the dominant contribution to the cross section in the low-$\mhl$ scenario and the scenario of the favored region are coming from the pair of the charged Higgs couplings to neutral CP-even Higgses. Thus, the process itself is useful to determine and confirm the couplings $c_{H^+H^-h^0}$ and $c_{H^+H^-H^0}$.


\section{Acknowledgement}
The numerical calculations reported in this paper were partially performed at TUBITAK ULAKBIM, High Performance and Grid Computing Center (\texttt{TRUBA} resources) and also at the computing resource of \texttt{fencluster} (Faculty of Science, Ege University). Ege University supports this work, the project number 17-FEN-054.

\bibliography{template-8s_revtex}


\end{document}